\documentclass[12pt]{article}

\usepackage[utf8]{inputenc}
\usepackage[T1]{fontenc}
\usepackage{color}
\usepackage{amsmath}
\usepackage{amsfonts}
\usepackage{amssymb}
\usepackage{caption}
\usepackage{graphicx}
\usepackage{slashed}            % for slashed characters in math mode
\usepackage{subcaption}
\usepackage{xspace}				% For spacing after commands
\usepackage{cite}
\usepackage{bm}
\let\originalleft\left
\let\originalright\right
\renewcommand{\left}{\mathopen{}\mathclose\bgroup\originalleft}
\renewcommand{\right}{\aftergroup\egroup\originalright}

\usepackage[english]{babel}
\usepackage{fancyhdr}
\usepackage{amsmath}
\usepackage{amssymb}
\usepackage{amsfonts}
\usepackage{psfrag}
\newcommand{\arXiv}[2]{\href{http://arxiv.org/abs/#1}{{\tt [arXiv:#1~[#2]]}}}
\newcommand{\arXivold}[1]{\href{http://arxiv.org/abs/#1}{{\tt [arXiv:#1]}}}

\renewcommand{\tilde}{\widetilde} % dinky tildes look silly

\newcommand{\beq}{\begin{eqnarray}}
\newcommand{\eeq}{\end{eqnarray}}

\newcommand{\bag}{\begin{align}}
\newcommand{\eag}{\end{align}}

\usepackage[hypertexnames=false]{hyperref}		% kills links in index (if exists)

\numberwithin{equation}{section}

\begin{document}
\begin{titlepage}

\begin{center} %TITLE HERE
{\huge \bf New Horizons in the Holographic \\ \vspace{0.1in} Conformal Phase Transition} 
\end{center}
\vspace*{0.4cm} 
\begin{center} % AUTHORS HERE
{\bf \  Cem Er\"{o}ncel$^a$, Jay Hubisz$^b$, Seung J. Lee$^c$,\\
Gabriele Rigo$^d$, Bharath Sambasivam$^b$}
\end{center}
%% PLACES HERE
\begin{center}
$^a$ {\it Istanbul Technical University, Department of Physics, 34469 Maslak, Istanbul, Turkey}\\
$^b$ {\it Department of Physics, Syracuse University, Syracuse, NY  13244}\\
$^c$ {\it Department of Physics, Korea University, Seoul 136-713, Korea}\\
$^d$ {\it Université Paris-Saclay, CNRS, CEA, Institut de Physique Théorique,\\
 91191, Gif-sur-Yvette, France\\}
\vspace*{0.6cm}
%% EMAILS HERE
{\tt  
\href{mailto:cem.eroncel@itu.edu.tr}{cem.eroncel@itu.edu.tr},
\href{mailto:jhubisz@syr.edu}{jhubisz@syr.edu},  
\href{mailto:sjjlee@korea.ac.kr}{sjjlee@korea.ac.kr},\\
\href{mailto:gabriele.rigo@ipht.fr}{gabriele.rigo@ipht.fr},
\href{mailto:bsambasi@syr.edu}{bsambasi@syr.edu}}
\end{center}
\vglue 0.3truecm
\begin{abstract}
We describe 5D dynamical cosmological solutions of the stabilized holographic dilaton and their role in completion of the conformal phase transition.  This analysis corresponds, via the AdS/CFT dictionary, to a study of out-of-equilibrium dynamics where trajectories of the dilaton do not depend solely on thermodynamic quantities in the early universe, but have sensitivity also to initial conditions.  Unlike the well-studied thermal transition, which requires quantum tunneling of an infrared brane through the surface of an AdS-Schwarzschild horizon, our approach instead invokes an early epoch in which the cosmology is fully 5-dimensional, with highly relativistic brane motion and with Rindler horizons obscuring the infrared brane at early times.   In this context, we demonstrate the existence of a large class of natural initial conditions that seed trajectories where the brane simply passes through the Rindler horizon and into the basin of attraction of the stabilized dilaton potential.  This corresponds to successful completion of the phase transition without sacrificing perturbativity of the 5D theory.

\end{abstract}
\end{titlepage}

\setcounter{equation}{0}
\setcounter{footnote}{0}
\setcounter{section}{0}

%%%%%%%%%%%%%%%%%%%%%%%%%%%%%%%%%%%%%%%%%%%%%%%%%%%%%%%%%%%%%%%%%%%%%%%%%%%%%%%%%%%%
%%%%%%%%%%%%%%%%%%%%%%%%%%%%%%%%%%%%%%%%%%%%%%%%%%%%%%%%%%%%%%%%%%%%%%%%%%%%%%%%%%%%

\section{Introduction}
Randall-Sundrum (RS) I models \cite{Randall:1999ee} are an attractive geometric solution to the problem of naturally creating large hierarchies of scale in physics.  They are built on 5D anti-de Sitter (AdS) space, and according to the AdS/CFT correspondence~\cite{Maldacena:1997re,Gubser:1998bc,Witten:1998qj}, they may be interpreted as 5D duals to strongly coupled 4D conformal theories where the conformal symmetry is spontaneously broken~\cite{Arkani-Hamed:2000ijo,Rattazzi:2000hs}.  Stabilization of these geometries is crucial for obtaining realistic low energy phenomenology as well as standard late-time 4D cosmology.

The issue of stabilization is clear in the low energy effective field theory (EFT) of spontaneously broken conformal symmetry.  When conformal invariance is broken at scale $\bar{f}$, there is a single Goldstone boson, the dilaton, that non-linearly realizes conformal invariance in the infrared (IR), undergoing a field shift under conformal transformations. An arbitrary EFT of the dilaton can be constructed by considering all operators consistent with the symmetries.  It contains a tower of higher dimensional operators whose coefficients are determined by details of the parent CFT, and also contains a conformally invariant non-derivative term, the dilaton quartic:
\beq
S = \int d^4x e^{-4 \tau} \left[ \frac{1}{2} \bar{f}^2 e^{2\tau} (\partial \tau)^2 - \lambda + 2 a e^{4\tau} \left( \partial \tau \right)^4 + \mathcal{O} ( \partial^6 ) \right].
\label{eq:dilac}
\eeq
A non-vanishing quartic coupling destabilizes the dilaton, meaning that there are no static solutions that spontaneously break the conformal symmetry unless $\lambda = 0$.  Generically, the field $\tau$ will evolve with time, and the effective vacuum expectation value (vev), $f = \bar{f} e^{-\tau}$ will move either towards zero or infinity depending on the sign of $\lambda$.
The first higher derivative term is closely related to the famous $a$-anomaly coefficient~\cite{Komargodski:2011vj}.  This term and the other higher derivative operators play a vital role if the initial conditions and/or dynamics inject considerable kinetic energy into the dilaton field.

In the dual 5D theory, the dilaton manifests as perturbations of a domain wall or brane that spontaneously breaks the isometries of AdS space.  The spacetime in RS models is characterized by a background metric
\beq
ds^2 = e^{-2 A(y)} dx_4^2 - dy^2,
\eeq
with $A= ky$ where $k$ is the curvature of AdS.  Very roughly, the holographic dual to the dilaton can be thought of as the scale factor, $A$, evaluated on a brane embedded as a function $y_\text{brane}(x)$.  It has recently been shown that the form of the action for this brane agrees with Eq.~(\ref{eq:dilac}) and also reproduces the correct $a$-anomaly coefficient in the effective action~\cite{Csaki:2022htl}.  

Phenomenological RS models also include an ultraviolet (UV) cutoff brane that corresponds to an explicit breaking of conformal invariance at scale $\mu_0$ by sourcing an effective 4D gravity theory with a finite Planck scale, and contributing an effective cosmological constant term through its tension.  Without any underlying mechanism in place, the tensions of both of the two branes must be exquisitely tuned against the bulk cosmological constant of AdS space to reproduce 4D flat space at long distances.  The two tunings of the IR and UV brane respectively create a flat direction (setting $\lambda = 0$) for the holographic dilaton and set the effective cosmological constant term to zero~\cite{Csaki:1999mp,Tanaka:2000er,Csaki:2000zn,Goldberger:1999un}.    

The Goldberger-Wise (GW) stabilization mechanism~\cite{Goldberger:1999un, Goldberger:1999uk} is the most-studied method of dynamically fixing the inter-brane distance in the RS I geometry.  In these models, a bulk scalar field obtains a coordinate dependent vacuum expectation value and its gravitational backreaction onto the geometry deforms the background away from AdS, effectively promoting the brane mistunes to dynamically varying quantities.  In this manner, the tuning of one of the brane tensions becomes unnecessary, and the remaining tuning is the one responsible for setting the effective cosmological constant to a tiny value.  Said in another way, the holographic dilaton is no longer an exact Goldstone boson in this background, whose dual corresponds to a near-marginal deformation of the CFT that leads to an IR gap through dimensional transmutation.   The dynamically generated  potential can stabilize the brane separation at distances that preserve these models as solutions to the hierarchy problem, fixing the scale of extra dimensional physics (or for restoration of conformal dynamics) to $\bar{f} \ll \mu_0$.

Despite the success for RS models in model building around the electroweak and flavor sectors, the simplest models of stabilization encounter strong constraints from early universe cosmology.  One of the primary issues is that of the conformal phase transition.  If one assumes the dual CFT to be in equilibrium with the thermal plasma after the big bang, then the system is in a hot conformal phase at temperatures $T \gg \bar{f}$.  From the perspective of the 5D theory, there is only the UV brane, and the finite temperature manifests as a deformation of AdS that terminates the geometry with an AdS-Schwarzschild horizon instead of an IR brane.  The conformal phase transition corresponds to nucleation of the IR brane through this horizon as the universe cools and the horizon recedes~\cite{Creminelli:2001th}.  The GW stabilization mechanism then relaxes the dilaton to the minimum of its stabilized potential.

The problem is that the phase transition is strongly first order, with nucleation and growth rates of ``true'' vacuum bubbles being too slow to compete with Hubble dilution in the false vacuum.  This nucleation rate per unit volume, $\Gamma/V \propto e^{-S_\text{bubble}}$, is exponentially sensitive to the inverse of the effective coupling constant of the 5D gravity theory, and so the perturbativity requirement is in tension with the necessity of having a successful phase transition.  In the dual picture, the rate depends on the number of colors, $N$ in the dual conformal gauge theory, through a large $N$ expansion of the bubble action, and so larger values of $N$ are associated with a slower tunneling rate.  See~\cite{Levi:2022bzt} for a recent analysis of these limitations, and ~\cite{Randall:2006py,Kaplan:2006yi,Hassanain:2007js,Bunk:2017fic,Dillon:2017ctw,vonHarling:2017yew,Megias:2018sxv,Agashe:2019lhy,Fujikura:2019oyi,Agashe:2020lfz,Agrawal:2021alq,Bruggisser:2022rdm,Baldes:2021aph,Bruggisser:2018mrt,Bruggisser:2018mus,Konstandin:2011dr,Baratella:2018pxi,Csaki22r}
for other discussions and various possible workarounds. 

A primary goal of our paper is to explore new \emph{out-of-equilibrium} (OOE) solutions to the problem of getting trapped in the eternally inflating solution at $f=0$.  By OOE, we mean that the dilaton vev is not simply a function of temperature, but rather has independent time dynamics, $f = f(T(t) ,t)$.  In the 5D picture this amounts to seeking cosmological solutions to the coupled scalar-Einstein equations that result in an unobstructed flow from some reasonable initial conditions at the big bang to the configuration that minimizes the potential, $f = \bar{f}$.

The complete solutions of the geometry far away from the minimum of the Goldberger-Wise potential are difficult to study due to the complications associated with taking into account the full time-domain dynamics of the Goldberger-Wise bulk scalar field coupled to 5D gravity.  However, as we will show in this paper, much of the dynamics can be inferred purely from consideration of effective \emph{dynamical} mistunes of the brane tensions associated with the GW solution that only slowly evolve with time in significant regions of the cosmological evolution, analogous to slow roll in standard 4D inflationary cosmology.  

There is a particular class of models where we have sufficient theoretical control to calculate the effective mistunes all the way back to the big bang. Our requirement for this to be the case is that the marginal deformation of the CFT remain marginal deep into the IR until RG flow is cut off by the Hubble length of the dS expansion in the $f=0$ configuration. This is somewhat similar to the requirements associated with obtaining a naturally light dilaton with mass $m \ll \bar{f}$~\cite{Bellazzini:2013fga,Coradeschi:2013gda}, where the $\beta$-function of the deformation must remain small over a very large range of scales.   In such models, the cosmological horizon present in the model with de Sitter (dS) branes effectively curtails uncontrollably large scalar backreaction effects, admitting a trustworthy calculation of the physics all the way to very large brane separation. 

Since we will be operating in a sort of slow-roll regime, with near-constant mistunes, our approach to understanding the OOE phase transition begins with a complete analysis of the bulk and brane cosmology with \emph{constant} mistunes of various sign.  While the 5D cosmology has been studied extensively (see, for example,~\cite{Kaloper:1999sm,Chacko:2001em,Binetruy:2001tc,Kumar:2018jxz,Karch:2020iit}), and the dual of RS II~\cite{Randall:1999vf} cosmology has been analyzed~\cite{Gubser:1999vj,Hebecker:2001nv,Marolf:2010tg}, an interpretation of the results in terms of the holographic dilaton action and in the language of the conformal phase transition has not, to our knowledge, been published.  

As we detail in Section~\ref{sec:UnstbCosmo}, all of these cosmologies begin with an epoch in which a UV brane localized observer sees a radiation dominated cosmology at early times.  In this era, the bulk geometry terminates in a Rindler horizon that recedes with time as the radiation dilutes via Hubble expansion.  If the UV brane tension mistune is non-zero, the cosmology eventually enters a regime where an effective cosmological constant determines the late time dynamics, which are either dS or AdS depending on the sign of the mistune.  The universe according to the UV brane observer then either inflates or crunches.  In the meantime, initial conditions and the IR brane mistune determine the trajectory and eventual fate of the IR brane.   Importantly, all trajectories traced back to early enough time begin with a large $\gamma$ factor corresponding to \emph{highly relativistic} motion of the IR brane away from the UV brane, with the IR brane out of sight beyond the Rindler horizon associated with the UV observer's proper time coordinates.  Some trajectories lose their velocity quickly enough so that the IR brane enters the causally accessible region for the UV observer, corresponding to a transition from the unbroken to the broken phase.

We emphasize that, in the early relativistic brane epoch, the entire tower of higher dimensional operators in the dilaton effective theory described in Eq.~(\ref{eq:dilac}) is important, as are effects associated with the sourcing of 4D gravity that explicitly break conformal symmetry, creating additional terms in the effective dilaton action.  A comprehensive study purely from the low energy standpoint would thus be quite limited since a naive truncation of the operator series would not correctly capture relativistic effects that are clear in the 5D picture.  It is the 5D gravity dual that affords us the power to resum the tower of operators into ones with relativistic constraints built in, corresponding in our study to solving the non-linear 5D Einstein equations and brane matching conditions.  Our dilaton effective action is thus accurate to leading order in the $1/N$ expansion and scalar backreaction effects, with additional $1/N$ suppressed corrections entering as higher dimensional curvature operators in the bulk, and additional brane localized operators.

In the approximation of small backreaction of the stabilization mechanism on the background AdS geometry, much of the cosmological picture we have described survives.  Slow rolling of the effective mistunes in various regions of the phase space for the holographic dilaton mean that at any point on a trajectory (so long as slow roll conditions are satisfied), we can read off an effective approximate \emph{first order} equation of motion for the brane, much as the slow roll approximation in inflation reduces the order of the field equations.  Just as slow roll conditions in inflation tend to be violated as the universe exits the inflationary phase and the universe reheats, our slow-roll conditions become invalid as the holographic dilaton enters the basin of attraction associated with its potential.  For us, we do not consider this an issue, as we concern ourselves primarily with success or failure of the phase transition.  Certainly further study of phenomenological aspects of this model are merited, and would entail a description of how energy in the dilaton dissipates into excitations of Standard Model (SM) particles, how this process depends on initial conditions, and whether there are other cosmological observables that are sensitive to these dynamics.

Our primary conclusion is that there is a large class of initial conditions that lead, without obstruction, to a successful phase transition.  The IR brane passes through the Rindler horizon without incident, and proceeds rapidly towards the minimum of the effective potential.  It is only initial conditions with extremely high initial relativistic $\gamma$-factor that reach and fall into a metastable attractor at $f=0$.  This is due to the effects of the radiation in the early epoch that contribute to the dilaton effective potential.  The metastable attractor does not exist at early times, being washed out by the effects of the radiation dominated cosmology on the 5D bulk geometry.  Rather than having sensitivity to a tunneling rate from a thermal CFT phase to a broken phase, we instead are sensitive to unspecified dynamics that precede ours in the universe's history.  As such, it is difficult to construct a measure for what constitutes a ``natural'' set of initial conditions, though we offer an attempt at this based on the assumption of equipartition of energy into all CFT degrees of freedom  (including the dilaton itself) at a cutoff time determined by naive dimensional analysis.  Under this assumption, the dynamics lead to a successful transition for all but extremely large values of $N$.

The organization of this paper is as follows. In Section~\ref{sec:UnstbCosmo}, we review the background 5D cosmology (with additional details in Appendix~\ref{sec:AppendixA}), where the only energy content in the bulk is a cosmological constant.  In this section we derive, numerically solve, and analyze solutions to the equation of motion for the IR brane for various signs of the brane tension mistunes. In Section~\ref{sec:NatTraj}, we discuss what constitutes a reasonable set of initial conditions for the IR brane, given an assumption of equipartition of energy between the dual CFT degrees of freedom at early time, and discuss its scaling with the number of colors in the dual CFT. In Section~\ref{sec:solutions}, we derive the equations of motion for the coupled scalar-gravity system for mistuned brane tensions, and discuss a dual CFT interpretation of the background scalar solutions. Following that, in the same section, we explore the two classes of stationary points of the dilaton potential- ones where the 5D geometry is cut off by 2 branes, and ones where the geometry is cut off by a brane and a Rindler-type horizon. In Section~\ref{sec:radionspec}, we establish the (in)stability of the stationary points we have identified. In Section~\ref{sec:StabCosmo}, we develop an approximate way to model effects of a time-dependent stabilizing Goldberger-Wise field for the cosmology, inspired by slow-roll inflation, and provide slow-roll conditions (with additional details in Appendix~\ref{sec:AppendixB}) for the validity of our approximation.  In this section we also address the question of the completion of the phase transition in light of trajectories for the IR brane in the slow-roll regime.

%%%%%%%%%%%%%%%%%%%%%%%%%%%%%%%%%%%%%%%%%%%%%%%%%%%%%%%%%%%%%%%%%%%%%%%%%%%%%%%%%%%%
%%%%%%%%%%%%%%%%%%%%%%%%%%%%%%%%%%%%%%%%%%%%%%%%%%%%%%%%%%%%%%%%%%%%%%%%%%%%%%%%%%%%

\section{Unstabilized Dilaton Cosmology}
\label{sec:UnstbCosmo}

A great deal of intuition for the physics associated with the more realistic stabilized dilaton scenario can be gleaned from a complete analysis of the case where there are constant mistunes of the IR and UV brane tensions.  This is because in certain regimes of evolution, the effects of a stabilization mechanism on the cosmological dynamics can be approximated by a situation with slowly evolving mistunes of these tensions, with variation entering primarily through the motion of the IR brane.  

In this section, we thus review the derivation of the bulk and brane cosmology of the detuned Randall-Sundrum model, and describe in detail the various scenarios that arise.  We pay particular attention to aspects of this model that continue to apply when a stabilization mechanism is in place.  We also comment on the AdS/CFT dual of the 5D cosmology in terms of a dynamical dilaton field in a strongly coupled dual with approximate conformal invariance.  In Section~\ref{sec:StabCosmo}, we discuss the cosmology of the stabilized system of branes.

We start with the following ansatz, where the UV brane embedding is flat, and the metric is gaussian normal with respect to an extra dimensional coordinate, $y$, which is normal to the UV brane:
\beq\label{eq:cosmoansatz}
ds^2 = n^2(y,t) dt^2 - a^2(y,t) d\vec{x}_3^2 - dy^2.
\eeq
The 5D geometry is compactified, with the UV brane at $y  = 0$, and an IR brane embedded according to the function $y = R(t)$.
In this work, we presume spatial curvature is absent.  We take as a boundary condition $n(y_0)$ = 1, so that global time is defined as the proper time for an observer constrained to the UV brane.  Using the AdS/CFT dictionary, such an observer is built out of fundamental particles external to the CFT, and potentially coupling weakly to it.

In a 5D space with only cosmological constant and tensions of the branes, the action is given by
\beq
S = \int d^5 x \sqrt{g} \left(-\Lambda_5 - \frac{1}{2 \kappa^2} \mathcal{R} \right) - \int d^4 \xi_0 \sqrt{G^0} T_0  - \int d^4 \xi_1 \sqrt{G^1} T_1,
\eeq 
where $\mathcal{R}$ is the Ricci scalar, and $G^{0,1}_{ab}$ are the pullback metrics for the UV and IR brane respectively.  The 5D Planck scale is given by the relation $\kappa^2 =1/ (2 M_5^{3})$.  The coordinates are chosen such that the embedding of the UV brane is trivial:  $G^0_{ab} = g(y=0)_{\mu\nu} \delta^\mu_a \delta^\nu_b $.  The same \emph{cannot} be done for the IR brane while maintaining the metric ansatz above, thus non-trivial motion of the IR brane is encoded in the functions relating the brane coordinates $\xi_1$ to the bulk coordinates.  Taking $\Lambda_5 = -\frac{6k^2}{\kappa^2}$, where $k$ is the curvature of the bulk AdS space, we write the IR and UV brane tensions in terms of dimensionless mistunes, $\delta_{0,1}$, as $T_0 = \frac{6k}{\kappa^2} (1+ \delta_0)$, and  $T_1 = - \frac{6k}{\kappa^2} (1- \delta_1)$. A calculation of the effective action under the assumption of a static geometry gives a form of effective potential for the dilaton, $R$~\cite{Bellazzini:2013fga,Eroncel:2019zev}:
\beq
V_\mathrm{eff} = \frac{6k}{\kappa^2} \left[ \delta_0 + e^{-4kR} \delta_1 \right].
\eeq
We thus refer to $\delta_0$ as the cosmological constant term, and $\delta_1$ as the dilaton quartic.  Importantly, there will be a rich interplay between these during the cosmological evolution of the full 5D theory, during which the dynamics are best described in the full extra dimensional theory, and where this effective potential gives only a coarse picture of the physics.  Working in terms of these dimensionless mistunes has the advantage that the factors of $\kappa^2$ cancel in the Einstein equations.  The solutions to the bulk geometry then depend only on the fractional tuning of the branes against the bulk cosmological constant.  
%This is also important when considering the $N$ counting for reasonable initial conditions for brane trajectories, where we employ the usual relation between the 5D gravity scale and the $N$ of the dual CFT:  $\kappa^2 k^3 = \frac{8 \pi^2}{N^2}$.  The fractional mistune parameters do not have any scaling with $N$~\cite{Chacko:2013dra}.
%\footnote{In the presence of stabilization, they get an $N$ scaling entering through $\kappa$.}

The calculation of the geometry as a function of $y$ and $t$ is reviewed in Appendix~\ref{sec:AppendixA}.  An important result is that in these coordinates, the bulk geometry does not depend on the IR brane mistune, only on the UV brane mistune.  Solving the Einstein equations for a UV mistune $\delta_0$ yields a result for the Hubble expansion rate on the UV brane:
\beq
\frac{1}{k^2} H^2 = \frac{4\bar{\lambda}}{\bar{a}^4} + \delta_0 \left( 2 + \delta_0 \right),
\label{eq:HubbleUV}
\eeq
where we have defined $\bar{a} = a(y=0,t)$.  There is a ``dark radiation'' term, proportional to an integration constant $\bar{\lambda}$ that arises in solving the Einstein equations~\cite{Csaki:1999mp}.  There is also a cosmological constant term corresponding to the mistune of the UV brane tension, $\delta_0$.  Note that with a negative sign cosmological constant term, the UV brane cosmology will undergo a ``crunch'' in finite time.  Interestingly, the Hubble rate is not as simply related to the effective potential for the theory as it would be in 4D, with the IR brane itself not contributing to the cosmology whatsoever, despite it being a field with a potential term (associated with the IR mistune $\delta_1$) that should gravitate.   In situations where the cosmology is approximately 4D, such that the Hubble rate is nearly constant across the bulk, additional relations are enforced between the UV and IR brane parameters that restore the naive expectation where the effective potential contributes as a source for the cosmology.  In the general 5D case, however, $\delta_0$ and $\delta_1$ are truly independent parameters and the cosmology of a UV observer does not ``see'' the moving brane.\footnote{We are considering only the case of pure 5D gravity.  Other 5D degrees of freedom may serve as a medium for communication across the bulk and change this prediction.  An interesting example would be a scalar field with a bulk shift symmetry, which is expected to preserve the dual conformal invariance up to scales associated with the UV brane, and whose vev can give a ``soft-wall'' shutoff of the geometry in place of the IR brane.  This soft wall dilaton vev would have some overlap with the UV brane, and presumably affect the cosmology there.} Also note that while the cosmological constant can be negative, it can not be more negative than the 5D curvature squared, which occurs when the UV brane tension is zero.  In the remaining discussion, we consider small mistunes of the brane tensions, so that the sign of $\delta_0$ determines the sign of the contribution to the Hubble equation.

The full solutions for the metric functions $a(y,t)$ and $n(y,t)$ are given in Appendix~\ref{sec:AppendixA}. The bulk spacetime has a Rindler horizon (or at times two) whose position depends on time. At early times, when the radiation term in Eq.~(\ref{eq:HubbleUV}) dominates, there is a horizon close to the UV brane. As time passes, and the radiation diminishes with the expansion of the universe (as seen by the UV brane observer), this horizon recedes. If $\delta_0$ is positive, the epoch of radiation domination eventually yields to de Sitter expansion, while if it is negative, the universe begins AdS crunching, and the radiation term again becomes important as the universe closes. The presence of horizons indicates the existence of infrared scales that are associated with temperatures throughout the cosmological evolution. This is the temperature associated with the dark radiation, or, additionally, due to a de Sitter temperature if the vacuum energy term is positive. Through the AdS/CFT correspondence, this bulk 5D side-effect of cosmological evolution corresponds to the CFT's response to the breaking of conformal invariance through the non-vanishing Hubble rate; this 4D quantum effect of Hawking radiation is encapsulated by classical physics in the 5D picture.

The motion of the IR brane is obtained by solving the matching conditions for the brane-bulk geometry, also discussed in Appendix~\ref{sec:AppendixA}, which yields a first order differential equation for $R$~\cite{Binetruy:2001tc}:
\beq
\frac{\frac{\dot{R}}{n} \frac{\dot{a}}{an} + \frac{a'}{a}}{\sqrt{1- \left(\frac{\dot{R}}{n}\right)^2} } = -1 + \delta_1,
\label{eq:IRbraneEq}
\eeq
where $a$ and $n$ are evaluated as functions of $y=R(t)$ and $t$.  This junction condition can be viewed as an integral of a second-order equation of motion, and corresponds to conservation of energy of the brane-geometry system.  A second-order equation can also be derived from the junction conditions, and this can be utilized to put the evolution equation in a more familiar form.  There are different ways to do this that give different (but equivalent) forms of this equation - for our purposes, we plug the junction conditions in the $yy$ Einstein equation and find:
\beq
\ddot{R} + \left[ \left(3 - \frac{1}{n^2} \frac{\partial V}{\partial R} +\tilde{f}(R,\beta) \right) \frac{\dot{a}}{a}- \frac{\dot{n}}{n} \right] \dot{R} + \frac{\partial V}{\partial R} =0.
\label{eq:2ndorder}
\eeq
We have packaged all relativistic corrections into the function $\tilde{f}(R,\beta)$, where $\beta \equiv \dot{R}/n$. This second order equation is just the expected one for the evolution of a scalar field experiencing a form of Hubble friction and having an effective potential.
The trajectory for the brane in the non-relativistic regime is thus governed by two features of the 5D cosmological dynamics:  the friction term, and an effective potential gradient.

The effective potential gradient term is given by 
\beq \label{eq:lindil0}
	\frac{\partial V}{\partial R} = - n^2 \left[ 4 \delta_1 + \frac{1}{n} \Dot{\Tilde{H}} + 2 \Tilde{H}^2 \right] \approx -4 (\delta_1 e^{-2kR} + \delta_0) + {\mathcal{O}} (\left(\mu_{\text{IR}}/f\right)^8),
\eeq
where $\tilde{H}=\frac{\dot{a}}{an}$. The form of this equation is as expected from a spontaneously broken approximately conformal theory, gapped by a compensated mass term, and by other IR scales.\footnote{The terms we have neglected are higher order in the large $t$ expansion, and involve a mixture of IR scales such as the dark radiation temperature and the de Sitter temperature.  We have taken these to be hierarchically smaller than the dilaton vev, roughly related to the IR brane position by $f \sim e^{-kR}$.  In terms of the geometry, this limit corresponds to the IR brane being well inside of the horizon.} From the perspective of the effective theory of the dilaton, the first few terms (in a derivative expansion) of the non-linearly realized gapped CFT action are
\beq
S = \int d^4x \sqrt{g} \left[ \frac{1}{2}e^{-2kR} (\partial R)^2 - \frac{1}{2} m^2 e^{-2kR} - \lambda e^{-4kR} \right].
\eeq
The equation of motion for the background solution on an FRW background is
\beq
\ddot{R} + 3 H \dot{R} - m^2 - 4  \lambda e^{-2kR} = 0,
\eeq
where we have dropped the term at second order in the field fluctuations.  This leads us to identify $\delta_1$ with the quartic term, and $\delta_0$ with the mass gap.  In Appendix~\ref{sec:AppendixA}, we showed that the bulk cosmology is dictated by the value of $\delta_0$ and a 5D ``dark radiation''.  The dark radiation does not give a huge effect in the effective potential due to the fact that the dilaton couples to the trace of the energy momentum tensor which does not get contributions from radiation, which is classically conformal.  We finally note that the relativistic corrections that re-sum to the full equation of motion arise from the sequence of higher derivative operators in the theory, such as those arising from the conformal anomaly for the CFT~\cite{Csaki:2022htl}.  These must conspire to reproduce the relativistic constraints that are a natural output of the solutions to the Einstein equations with a sensible 5D gravity action.

\subsection{Characterizing the solutions}
\label{subsec:solutions}
\begin{figure}[h!]
\center{
\includegraphics[width=0.45\textwidth]{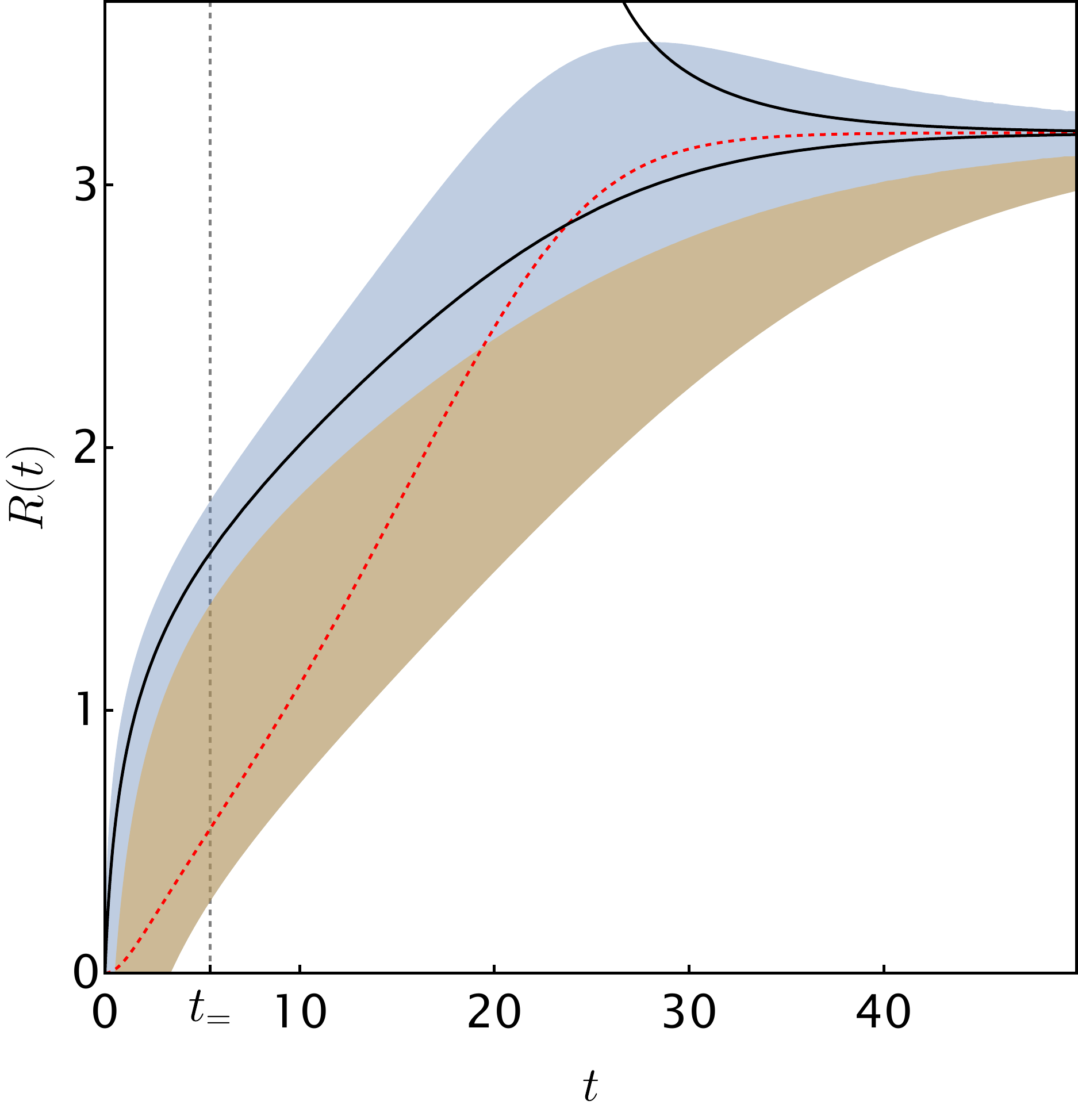}
\includegraphics[width=0.45\textwidth]{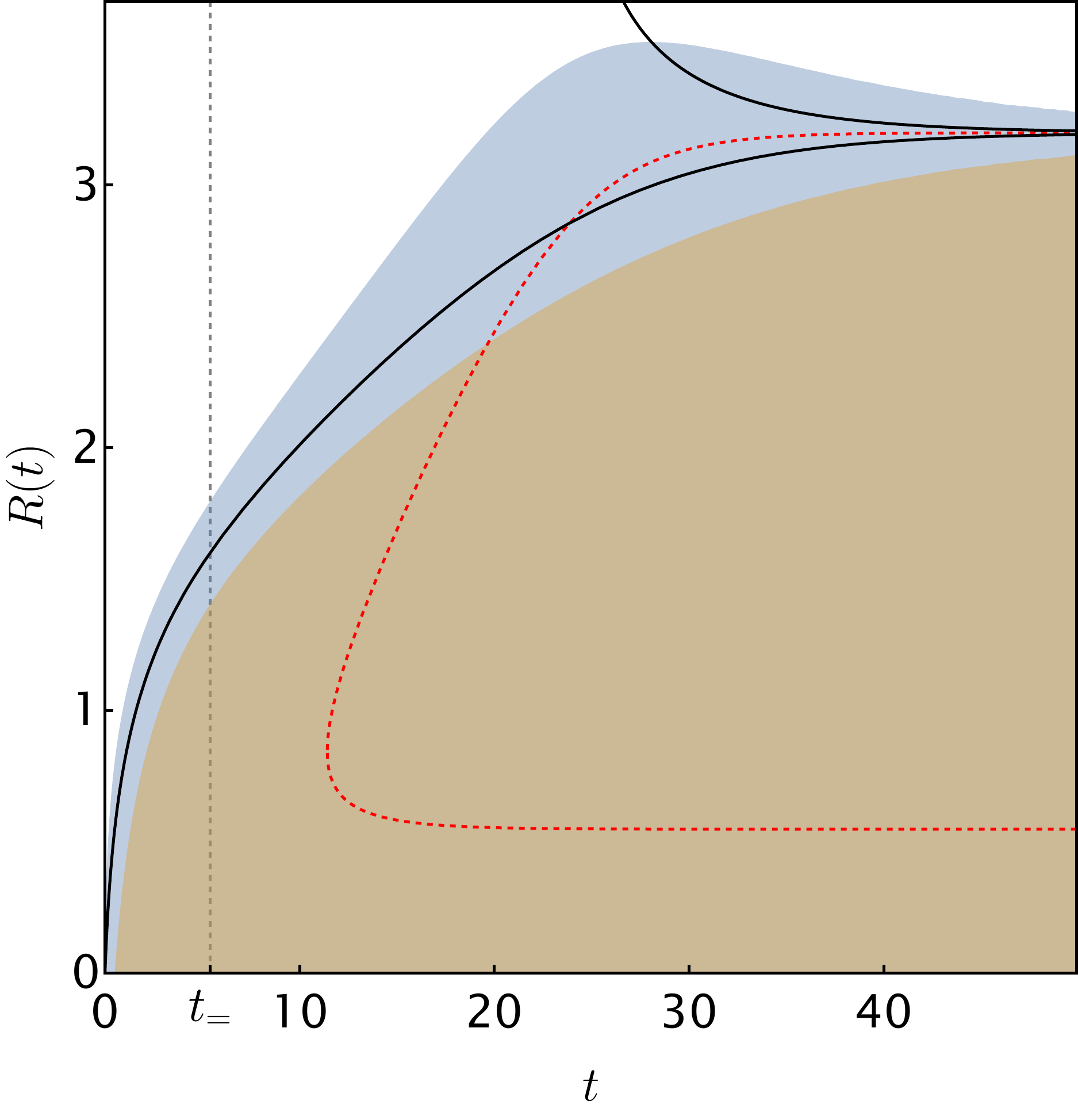}
\includegraphics[width=0.45\textwidth]{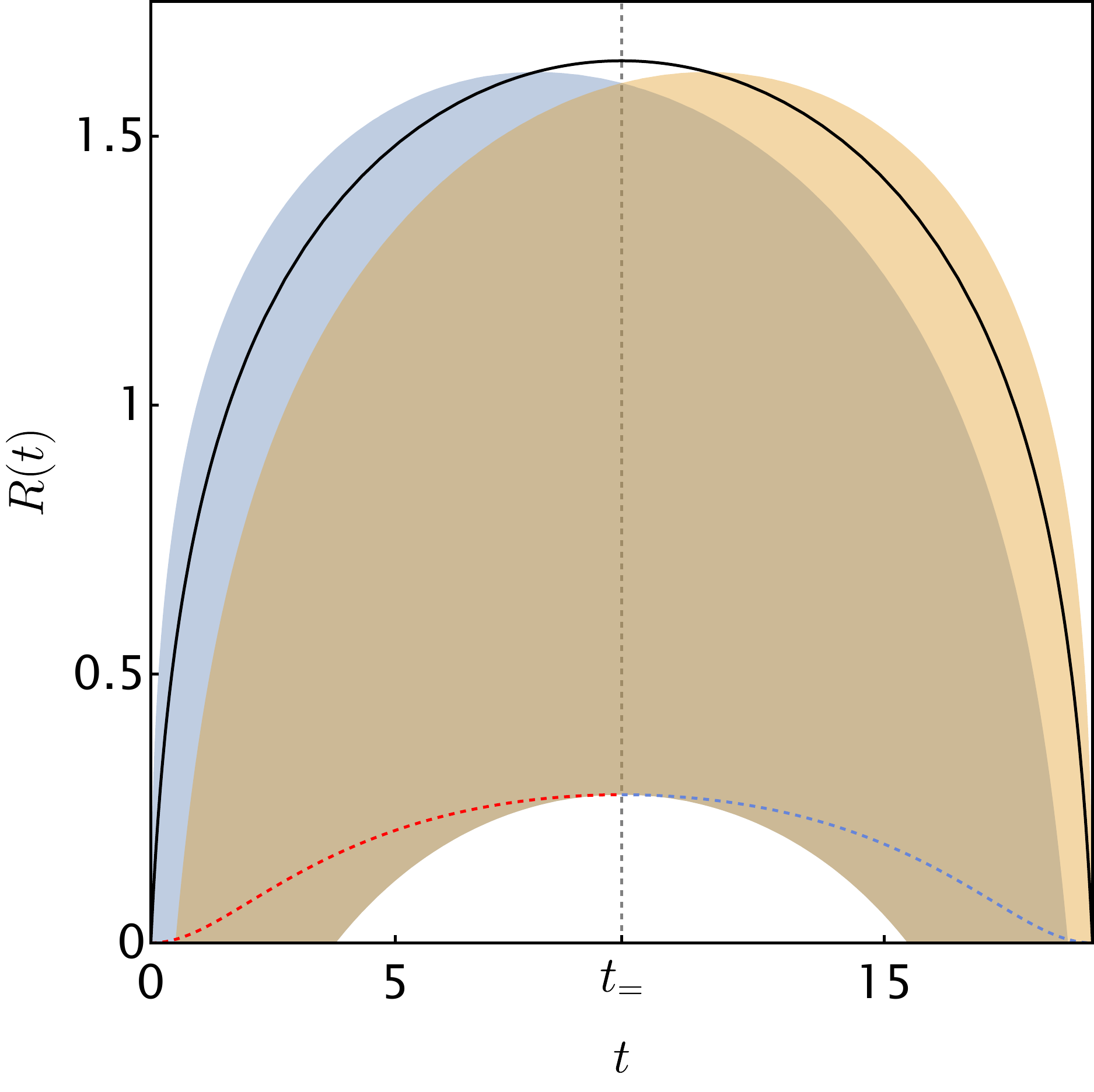}
\includegraphics[width=0.45\textwidth]{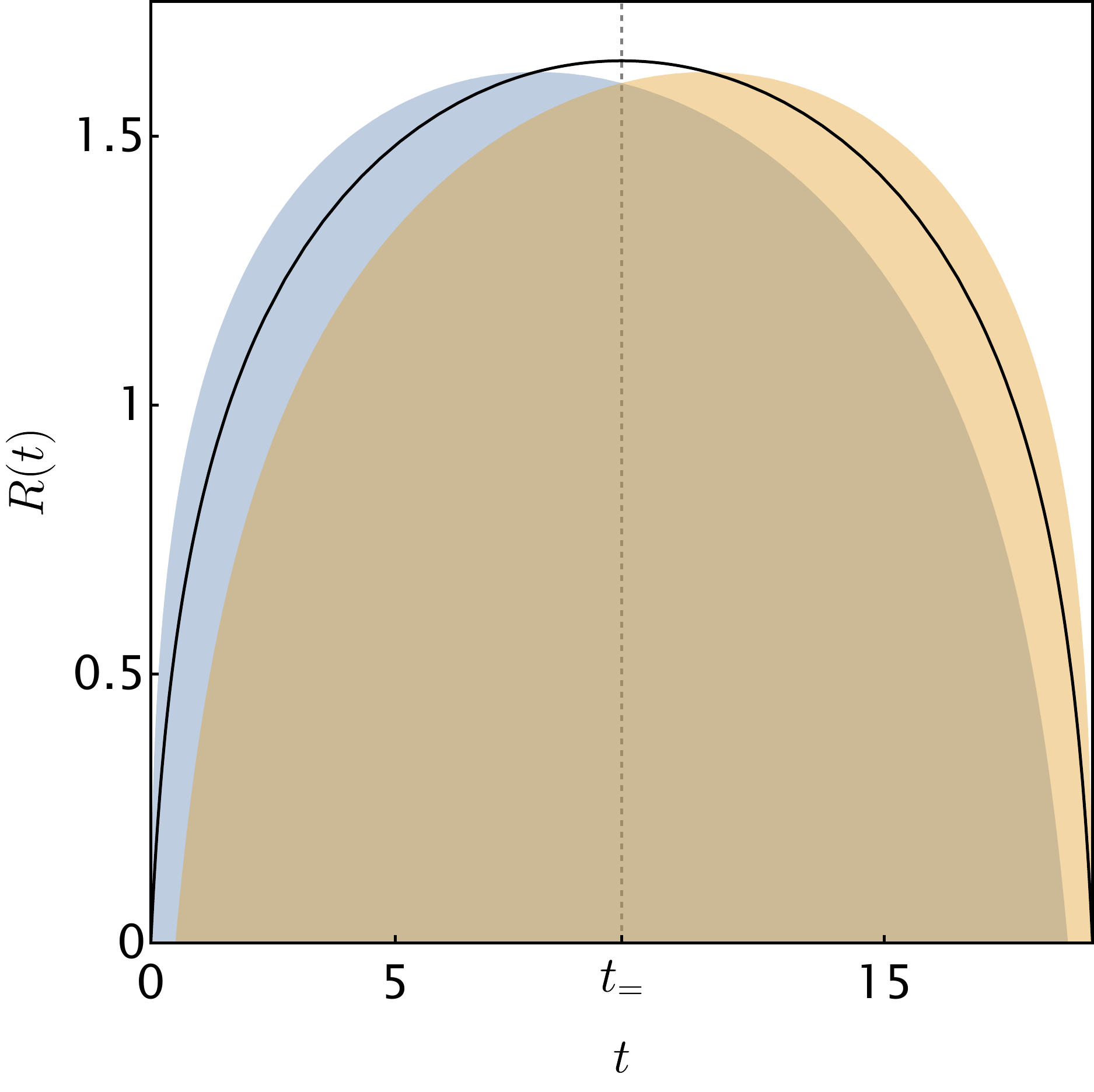}}
\caption{Allowed regions for the brane for various sign choices of $\delta_0$ and $\delta_1$ ($++,+-,-+,$ and $--$ in reading order).  The blue (orange) colored regions are the regions of validity for the first (second) branch for solutions to the quadratic equation obtained by squaring Eq.~(\ref{eq:IRbraneEq}).  Regions of overlap appear as the darker mix of blue and orange.  The red (blue) dashed lines correspond to stationary points $\beta = 0$ for the first (second) of the two branches.  The black lines correspond to the horizon(s). We have used  $|\delta_0| = 1/300$, and $|\delta_1| = 1/100$ for all plots.  We have indicated the time of radiation/vacuum energy equality in each plot, $t_=$.  In the case of negative $\delta_0$ in the last two panels, radiation always dominates, except at the single moment when $H = 0$ before the universe begins crunching.}
\label{fig:branches}
\end{figure}

Squaring Eq.~(\ref{eq:IRbraneEq}), and defining $\beta \equiv \dot{R}/n$, we see that there are two possible branches for the solutions for $\beta$ from which we can arrive at a first order equation $\dot{R} = \sigma(R,t)$.  However, it is possible that one or neither of the two solutions is valid/physical for a given value of $R$ and $t$.  Curiously, this implies that there are regions in the 5D space-time where the brane simply cannot exist.  One can understand this intuitively from energy conservation:  in part, the cosmology is driving the motion of the brane up its quartic potential, and for a given cosmology, the brane can not gain enough energy to make it past certain points.  In Figure~\ref{fig:branches}, we display the location of the horizons (zeros in the $n$ function, black solid lines) along with the physical regions for each of the two solutions to the quadratic equation for $\beta$ (colored in blue and orange).

Of particular interest are stationary points in the evolution  -- those where $\beta = 0$, corresponding to lines in the $R$ -- $t$ plane along which $a'/a = -1+\delta_1$.  When the brane crosses these lines, the moving brane experiences a turn-around.   In the large time limit, these lines correspond to stable or unstable fixed points for the trajectories -- stationary points in the effective dilaton potential.  The $\beta = 0$ lines are shown in Figure~\ref{fig:branches} as the dashed red (corresponding to the blue branch for $\beta$) or blue lines (corresponding to the orange branch).  These change with time for two reasons.  First, each point on the plot corresponds to a trajectory with different initial conditions where the brane has more or less initial relativistic energy, and will thus reach different points on its effective potential before turning around.  Second, the effective potential for the dilaton in this cosmological background is itself changing with time.  This is as expected for a CFT interpretation where the dilaton receives contributions to its effective potential from the scale invariant quartic, $\delta_1$, but also from explicit breaking of conformal invariance due to the cosmological constant term, $\delta_0$, and/or due to explicit breaking from to the temperature we would associate with the dark radiation.   We extract this temperature from the relation
\beq
H^2 = \frac{1}{3 m_\text{Planck}^2} \rho_R = \frac{4\bar{\lambda}}{\bar{a}^4} = \frac{1}{4t^2} \approx 4 k^2 e^{-4 y_H} \equiv \frac{4 \pi^4}{k^2} T^4.
\label{eq:hubbletemp}
\eeq
We have input the solution for the epoch of radiation domination, $\bar{a}^2 = 4 \sqrt{\bar{\lambda}} t$ and then re-expressed time in terms of the location of the horizon.  In the last statement, we have associated a temperature with the horizon location (see Appendix~\ref{sec:AppendixA}), $T = \frac{k}{\pi} e^{-ky_H}$.

The structure of the turn-around points is especially interesting in the second panel where the cosmological constant is positive, and the quartic is negative.  With a pure quartic term, the dilaton would always run to infinity, and there would be no CFT.  In the 5D picture, this means the IR brane would quickly collide with the UV brane, abruptly ending our 5D description of the cosmology.  The presence of the positive cosmological constant term changes this picture.  The cosmological constant term leads to a term in the potential that becomes important when $k y \gtrsim \log \delta_0$.  This term gaps the dilaton and gives rise to a metastable attractor solution where the IR brane asymptotically approaches the horizon induced by the dS evolution.  This is evident in Eq.~(\ref{eq:lindil0}), as there is, at late times, a zero in the potential gradient term when $R = \frac{1}{2} \log \frac{|\delta_1|}{\delta_0}$.

Importantly, only one of the two branches is ever valid near $t=0$, indicated as the blue regions in Figure~\ref{fig:branches}.  If the second branch is chosen, the IR brane will have ``emerged'' from the UV brane at some finite time $t_0 > 0$.  We do not trust such solutions below $t_0$, as the detailed physics giving rise to the branes must be taken into account when they are very close.  Because of this, we only consider solutions on the first branch which are valid to $t \approx 0$.\footnote{Of course, as with 4D cosmology, we must still cut off the trajectories at some finite $t_*$ corresponding to the regime where the effective ``temperature'' of the dark radiation term is larger than the cutoff scale, and where the kinetic energy of the brane is similarly large.}

\begin{figure}[h!]
\center{
\includegraphics[width=0.45\textwidth]{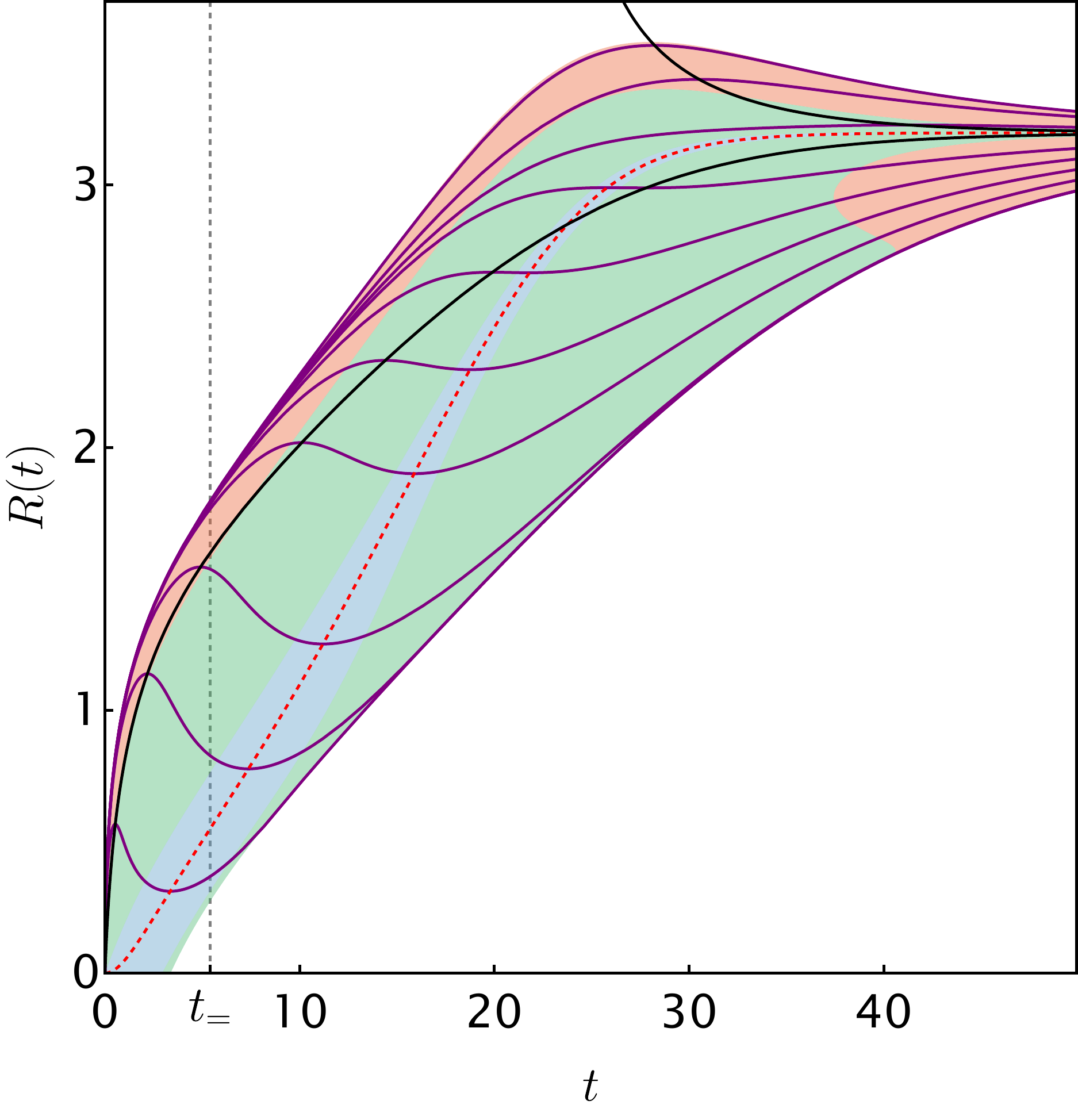}
\includegraphics[width=0.45\textwidth]{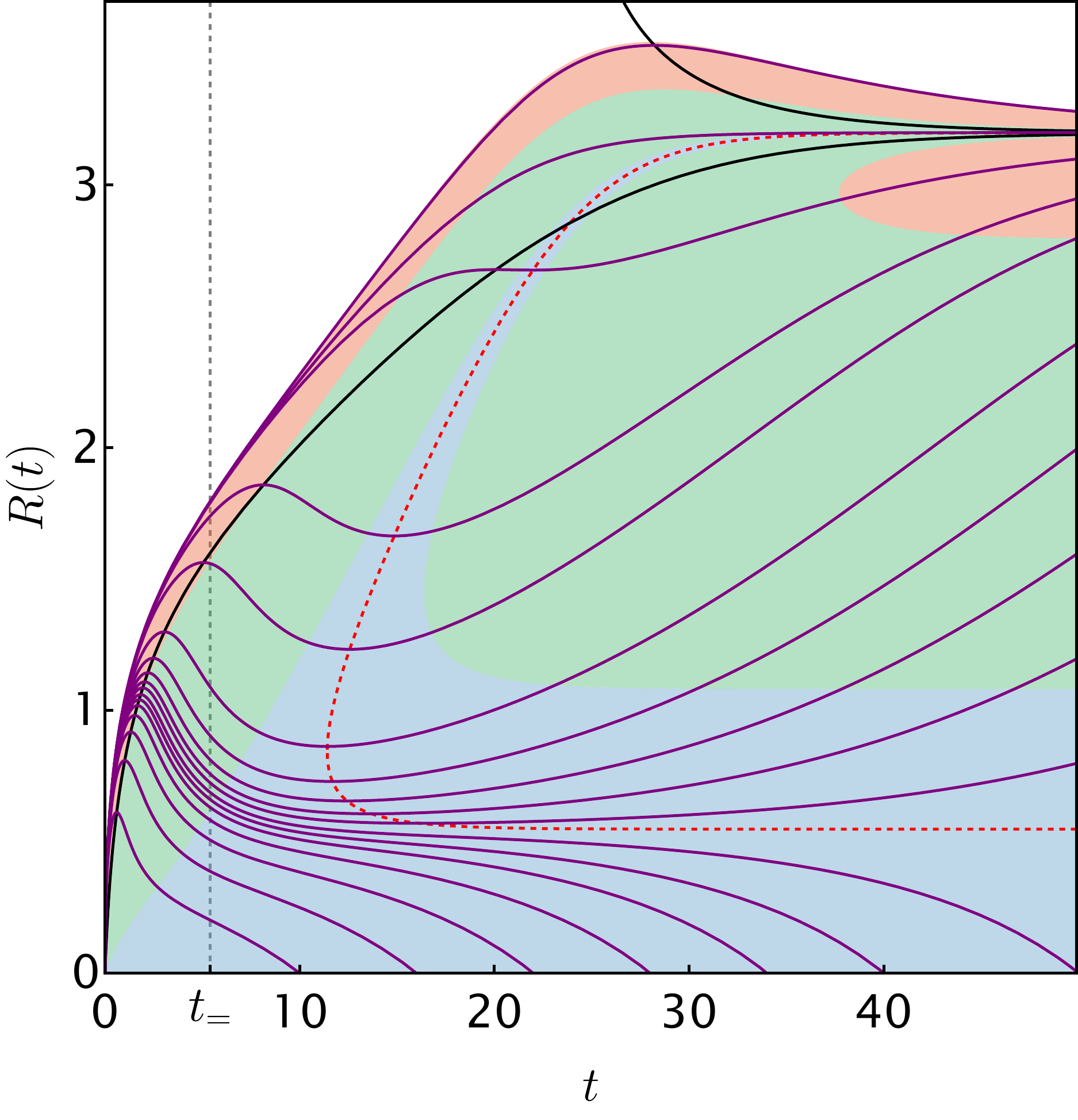}
\includegraphics[width=0.45\textwidth]{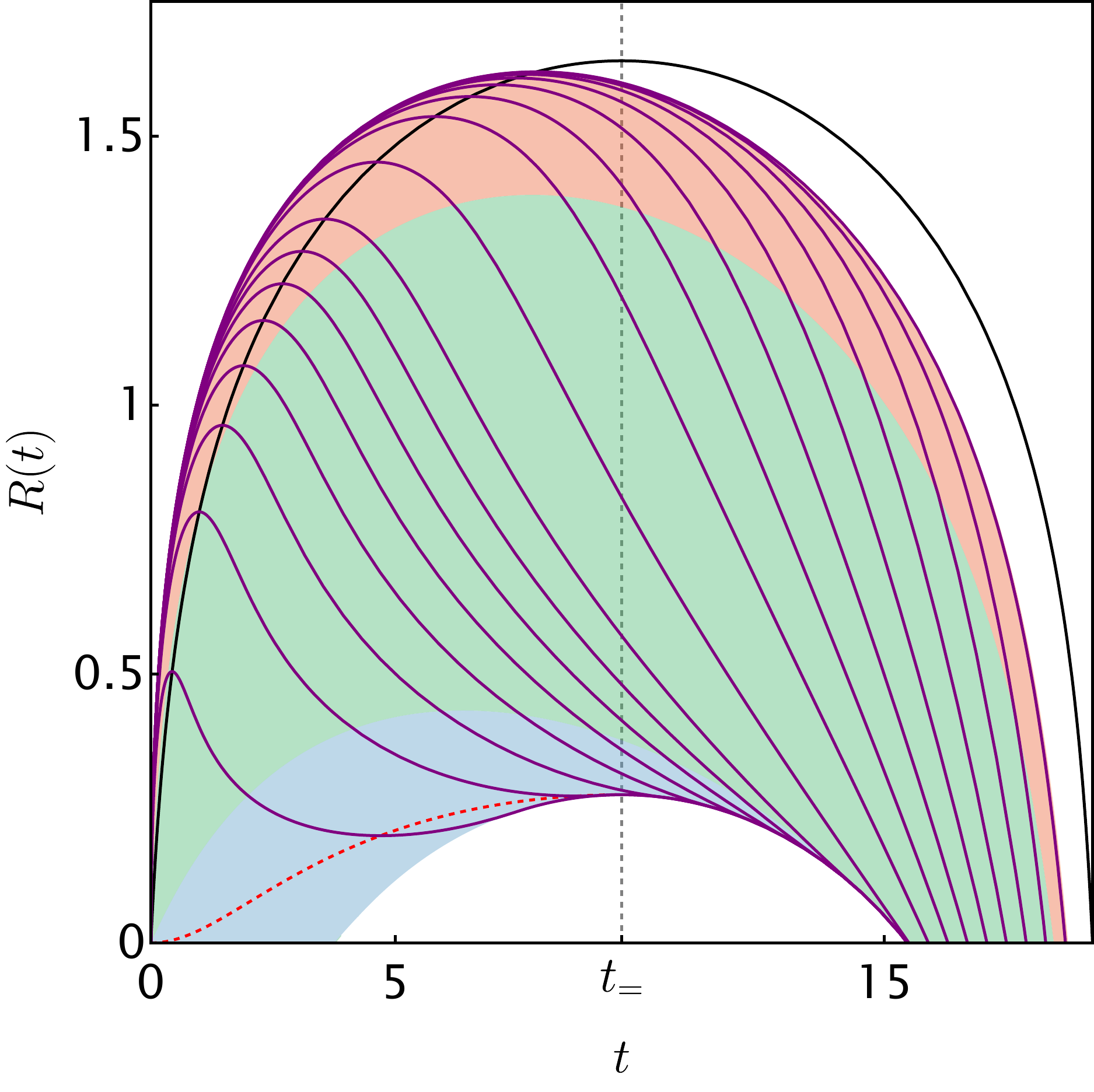}\hspace{.1in}
\includegraphics[width=0.45\textwidth]{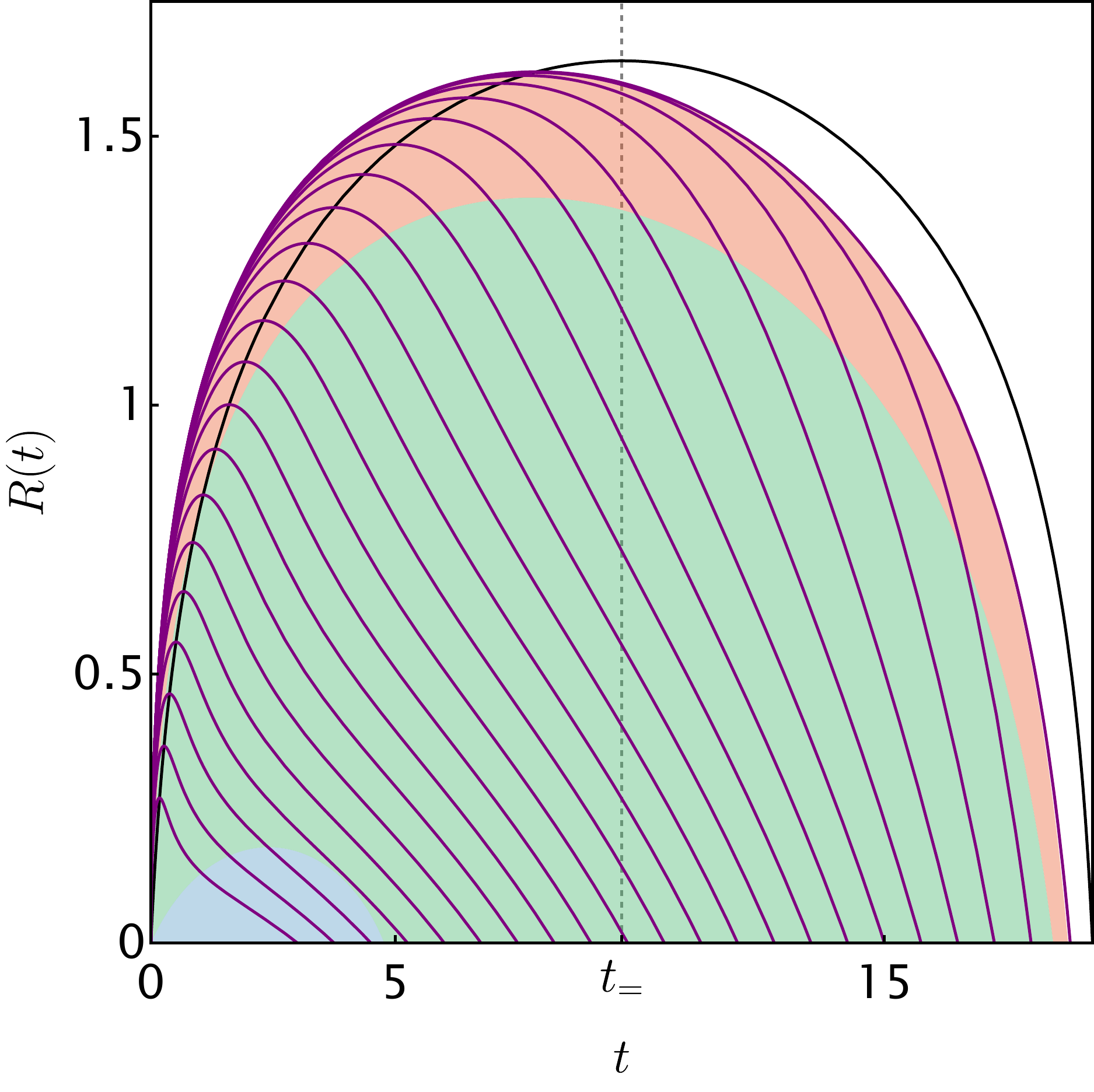}}
\caption{Here we display brane trajectories (purple) with a variety of initial conditions for various sign choices of $\delta_0$ and $\delta_1$ ($++,+-,-+,$ and $--$ in reading order). We have used here $|\delta_0| = 1/300$, and $|\delta_1| = 1/100$ for all plots, and used only the ``blue'' branches (see Figure~\ref{fig:branches}) of Eq.~(\ref{eq:IRbraneEq}) to obtain the brane velocity.   The horizon(s) are shown as black lines.  Turn-around points where the brane velocity is zero are shown as dashed red lines.  Contours indicate the relativistic velocity of the brane, with (red)/(green)/(blue) corresponding to $(1 > \beta \ge 0.9) / (0.9> \beta \ge 0.1) / (0.1 > \beta)$.}
\label{fig:ExTrajPlots}
\end{figure}

For a given choice of the cosmological constant and quartic ($\delta_{0,1}$) there are many trajectories, each with different initial conditions.  In Figure~\ref{fig:ExTrajPlots} we show these trajectories, assuming we have taken the branch of Eq.~(\ref{eq:IRbraneEq}) that allows one to extend solutions close to the big bang, at $t=0$.  Most importantly for the primary focus of this work, we see that the ultimate fate of the brane is determined by the initial conditions.  

In the cases of positive cosmological constant (the first and second panel) there are various possibilities for the late time position of the brane:  behind the horizon, approaching the horizon, or collapsed back to the UV brane.  If the initial conditions are very highly relativistic (for either sign of the quartic mistune, $\delta_1$), the IR brane crosses a second Rindler horizon that slowly merges with the first, and the IR brane never crosses into the region accessible by observers on the UV brane.  From the CFT point of view, there is no epoch in the cosmological evolution in which one can describe the CFT as having been spontaneously broken.  If the quartic mistune is positive (first panel), and the IR brane makes it into the visible region, the quartic eventually drives the IR brane back towards the first Rindler horizon, corresponding to $f \rightarrow 0$.  In this case, we would say that the CFT undergoes a transition into the broken phase at some critical temperature, but that the vev eventually evolves back towards zero, with the theory asymptoting to an unbroken CFT again at large times.  If the quartic mistune is negative (second panel), the least relativistic trajectories soon collapse back to the UV brane.  However, higher initial velocities for the brane can correspond to trajectories that get trapped in a metastable configuration and are then driven towards the Rindler horizon. This behavior is due to the explicit breaking of conformal invariance from the accelerated expansion.  The potential is no longer simply a quartic, but receives contributions from cosmological corrections.  If the brane is relativistic enough, it can make it into this basin of attraction for the trajectories created by backreaction effects on the 5D side, or quantum cosmology effects on the CFT side.

In the case of negative cosmological constant (the last two panels), it is curious to note that the motion of the brane spontaneously breaks the time-reflection symmetry of 4D crunching scenarios that arise in the case of a negative vacuum energy ($\delta_0<0$).  This reflection exchanges big bang for big crunch about the time of equality, $t_{=}$, or $H = 0$.  The choice of either of the two branches of Eq.~(\ref{eq:IRbraneEq}) amounts to a spontaneous breaking of this symmetry, or to a dynamical selection for the preferred direction for the arrow of time.  Our late-time description of the cosmology breaks down not at the big crunch singularity, but instead when the IR brane collides with the UV brane.  

Additional singular structure is evident for a class of trajectories in the case where $\delta_0 < 0$, and $\delta_1 > 0$, shown in the third panel.  All trajectories which cross the red dotted turnaround line before $t_=$ are focused through the single point that lies at the apex of the boundary of the allowed region.  This point is a singular point in the geodesic equation for the brane, and corresponds to the one point where both branches for the brane velocity coincide at $\beta = 0$.  
The trajectories cannot then be uniquely determined, since it is ambiguous which branch to use.  If one continues using the first branch, the focused family of geodesics all coincide, tracing out the boundary of the allowed region.  The past history of the brane cannot then be uniquely determined from its trajectory data once it has been focused.  If one switches branches, one must impose some additional constraint to uniquely continue the evolution past $t_=$.\footnote{While not of direct import for our main focus, it is amusing to note that imposing time-reflection symmetry about $t_=$ on this family of trajectories gives something like a ``dilaton bounce'' cosmology where the dilaton bounces at the half-cycle of the crunch cosmology.}

We now attempt to interpret the radiation density of Eq.~(\ref{eq:hubbletemp}) in terms of a strongly coupled conformal plasma in a dual large $N$ gauge theory.  The effective number of relativistic degrees of freedom in kinetic equilibrium, $g_*$, will be some numerical factor times the number of gluons in the dual gauge theory, $g_* \propto N^2 = \frac{8 \pi^2}{\kappa^2 k^3}$.  In large $N$ strongly coupled super Yang-Mills theory, this factor is famously $\frac{3}{4} \cdot 15$ where $15$ is the weakly coupled expectation, obtained from $n_b+\frac{7}{8} n_f$, and the suppression by $3/4$ corresponds to strong coupling effects~\cite{Gubser:1996de,Gubser:1998nz}.  Since the AdS/CFT correspondence is being applied in the large $N$ limit, the superpartners of the gravitational sector on the 5D side act only as weakly coupled spectators, and we thus expect that the counting of degrees of freedom in the dual theory at leading order in $N$ will be the same despite the lack of supersymmetry in the RS model.  This sort of ``hidden supersymmetry'' has been noted recently in the context of the $a$-anomaly associated with the dual to RS, and in 5D instanton calculations~\cite{Gherghetta:2021jnn,Csaki:2022htl}.  In equilibrium, the energy density will thus be 
\beq
\rho_R = \frac{\pi^2}{30} \frac{3}{4} \cdot 15 N^2 \frac{3}{4} \cdot 15 T^4 = \frac{3 \pi^2}{8} N^2 T^4= \frac{3}{\kappa^2 k^3} T^4.
\label{eq:SUSYrho}
\eeq
If we plug this into Eq.~(\ref{eq:hubbletemp}), with $m_\text{Pl}^2 = \frac{1}{4 \kappa^2 k}$, we find perfect agreement.

Finally, in the context of positive vacuum energy domination, we do find the expected relationship between the de Sitter temperature, $T_\text{dS} = \frac{H}{2\pi}$ and the horizon location.  The horizon location is given by $k y_H = \frac{1}{2} \log \frac{2+\delta_0}{\delta_0} \approx \frac{1}{2} \log \frac{2}{\delta_0}$ (taking the small $\delta_0$ limit), and we thus have
\beq
T_\text{dS} = \frac{H}{2\pi} \approx \frac{k}{\pi} \sqrt{\frac{\delta_0}{2}} =   \frac{k}{\pi} e^{-ky_H}.
\eeq

We see that dilaton cosmology is extremely rich, and that already in the simple case of constant mistunes there are many possible descriptions for the time evolution of the universe.  The fate of the universe and the dilaton depends not only on the 5D cosmological parameters $\delta_0$ and $\delta_1$, but also on the initial conditions of the dynamical dilaton.  Our main focus in this work is with the ultimate fate of the dilaton when one includes the dynamics of stabilization.  Under what conditions does the dilaton fall into the attractor solution associated with the global minimum of its potential? We can foresee the danger in the second panel of Figure~\ref{fig:ExTrajPlots} -- a metastable attractor that traps highly relativistic initial conditions and drives the dilaton vev to zero.

We then must then attempt to address the question of what constitutes a reasonable choice for the initial conditions for the dynamical dilaton.

%%%%%%%%%%%%%%%%%%%%%%%%%%%%%%%%%%%%%%%%%%%%%%%%%%%%%%%%%%%%%%%%%%%%%%%%%%%%%%%%%%%%
%%%%%%%%%%%%%%%%%%%%%%%%%%%%%%%%%%%%%%%%%%%%%%%%%%%%%%%%%%%%%%%%%%%%%%%%%%%%%%%%%%%%

\section{Initial Conditions}
\label{sec:NatTraj}
In this section, we address the question of what reasonable initial conditions for the IR brane might be. We strongly caution the reader that this is subject to details of the unspecified dynamics that gives rise to the geometric description we are studying, and that this analysis is meant only to provide a sample of how one might estimate the initial conditions for the brane motion.  For example, the cosmology we are describing might be dynamics that follows a period of inflation, and the precise details of how inflation exits and transfers energy into the system will determine what constitutes a natural set of initial conditions.

In this section, we suppose that the unspecified dynamics injecting energy into the system at some early cutoff time did not grossly discriminate between the various degrees of freedom in the theory, and that, in particular, they did not single out the dilaton.  In this case, at some early cutoff time $t_{*}$ during the epoch of dark radiation domination and small inter-brane separation, we might expect the partition of energy in the system to be roughly equal.  That is, the energy density of the brane at time $t_{*}$ is expected to be comparable to the energy density in one degree of freedom of the dark radiation:
\begin{equation}\label{eq:EqPartition}
	\rho_{\text{dilaton}} \sim \frac{1}{N^2}\rho_{R}\sim \frac{\pi^2}{30} T^4,
\end{equation}
where $N$ is the number of colors in the dual CFT, related to the 5D parameters through $\kappa^2 k^3 = \frac{8 \pi^2}{N^2}$, and we are ignoring $\mathcal{O}(1)$ factors.  This seems natural due to the fact that the brane is a single scalar modulus degree of freedom in the dual picture. We use this condition to estimate what natural initial conditions for the IR brane might look like. 

The first order equation for the motion of the IR brane in Eq.~(\ref{eq:IRbraneEq}) describes the possible initial conditions for the brane motion, and can be recast in the following form:
\begin{equation}\label{eq:ReqnHform}
	\frac{6}{\kappa^2}  \frac{a^3 n}{\bar{a}^3} \left[ \gamma \beta \tilde{H} + \gamma H_y - \frac{\kappa^2}{6} T_1 \right] = 0,
\end{equation}
where $\tilde{H}=\dot{a}/an$, $H_y=a'/a$, $\bar{a}$ is the scale factor evaluated on the UV brane, and $\gamma$ is the relativistic boost factor.  We have multiplied by the overall metric factors since with their addition, this equation reads like a non-canonical relativistic dispersion relation\footnote{On the CFT side, this novel dispersion relation should arise from consideration of the entire sequence of higher derivative operators in the effective theory for spontaneously broken conformal symmetry, and encodes dual CFT data such as the $a$-anomaly, which contributes at order $\partial^4$.} for the brane.  The first term corresponds to a momentum density, while the remaining two terms are the energy associated with a dilaton action of a DBI form:
\beq
S_\text{dilaton} = \int d^4 x \sqrt{g} f(R,t) \left[ \lambda(R,t) \sqrt{1- g^{\mu\nu} \partial_\mu R \partial_\nu R} - T_1 \right],
\eeq
with the identification $\lambda = H_y$, $f = \frac{a^3 n}{\bar{a}^3}$, and $g_{\mu\nu}$ being the FRW metric corresponding to the effective 4D cosmology on the UV brane.

One can check numerically that both $\tilde{H}, H_y \sim \mathcal{O}(k)$ along the trajectories, even at relatively early times.  This is particularly apparent in the vicinity of the global minimum of the dilaton potential, with vanishing vacuum energy- the Goldberger-Wise (GW) minimum, where $a\sim e^{-ky}$. Then, we can read off an estimate of the energy density of the brane from Eq.~(\ref{eq:ReqnHform}):
\begin{equation}
	\rho_{\text{dilaton}}\sim\frac{6}{\kappa^2}\gamma H_y \sim 6 \frac{k\gamma}{\kappa^2}.
\end{equation}
Next, we employ condition Eq.~(\ref{eq:EqPartition}) at an early time, $t_{*}\sim\Lambda^{-1}$, where 
\begin{equation}\label{eq:5DCutOff}
	\Lambda=\left(\frac{\sqrt{3}}{2N}\right)^{1/3}2\pi\, M_5
\end{equation}
is the cutoff for the 5D theory \cite{Chacko:2013dra}.  This gives
\begin{equation}
	\left| \frac{a^3 n}{\bar{a}^3} \right| \frac{6 k\gamma}{\kappa^2}=\frac{\pi^2}{30}T^4=\frac{k^2 H^2}{120\pi^2}= \frac{k^2}{120 \pi^2} \frac{4\bar{ \lambda}}{\bar{a}^2(t_*)}.
\end{equation}
In the second equality, we have employed Eq.~(\ref{eq:hubbletemp}) to write the temperature in terms of the Hubble rate.
Then, using $\bar{a}^2(t)\simeq4\sqrt{\bar{\lambda}}t$ during dark radiation domination (see Appendix~\ref{sec:AppendixA}), we obtain a characteristic value of $\gamma$ at the cutoff time:
\begin{equation}\label{eq:GammaNScalingEq}
	\left| \frac{a^3 n}{\bar{a}^3} \right| \gamma(t_*)=\frac{1}{720 \tau_0^2} \left( \frac{3 \pi^2}{2}\right)^{1/3} N^{-4/3},
\end{equation}
where we have used $t=t_{*}\sim\tau_0 \Lambda^{-1}$, and held $m_{\text{Planck}}^2=M_5^3/k$ constant under change of $N$ and $k$.  The variable $\tau_0 \equiv 1/10 -10$ is a factor designed to account for $O(1)$ factors in the estimate.  The initial condition is then set at $t_*$ by feeding this into the first order equation for the brane, and determining which trajectory intersects with the solution.

\begin{figure}[t!]
\center{
\includegraphics[width=0.6\textwidth]{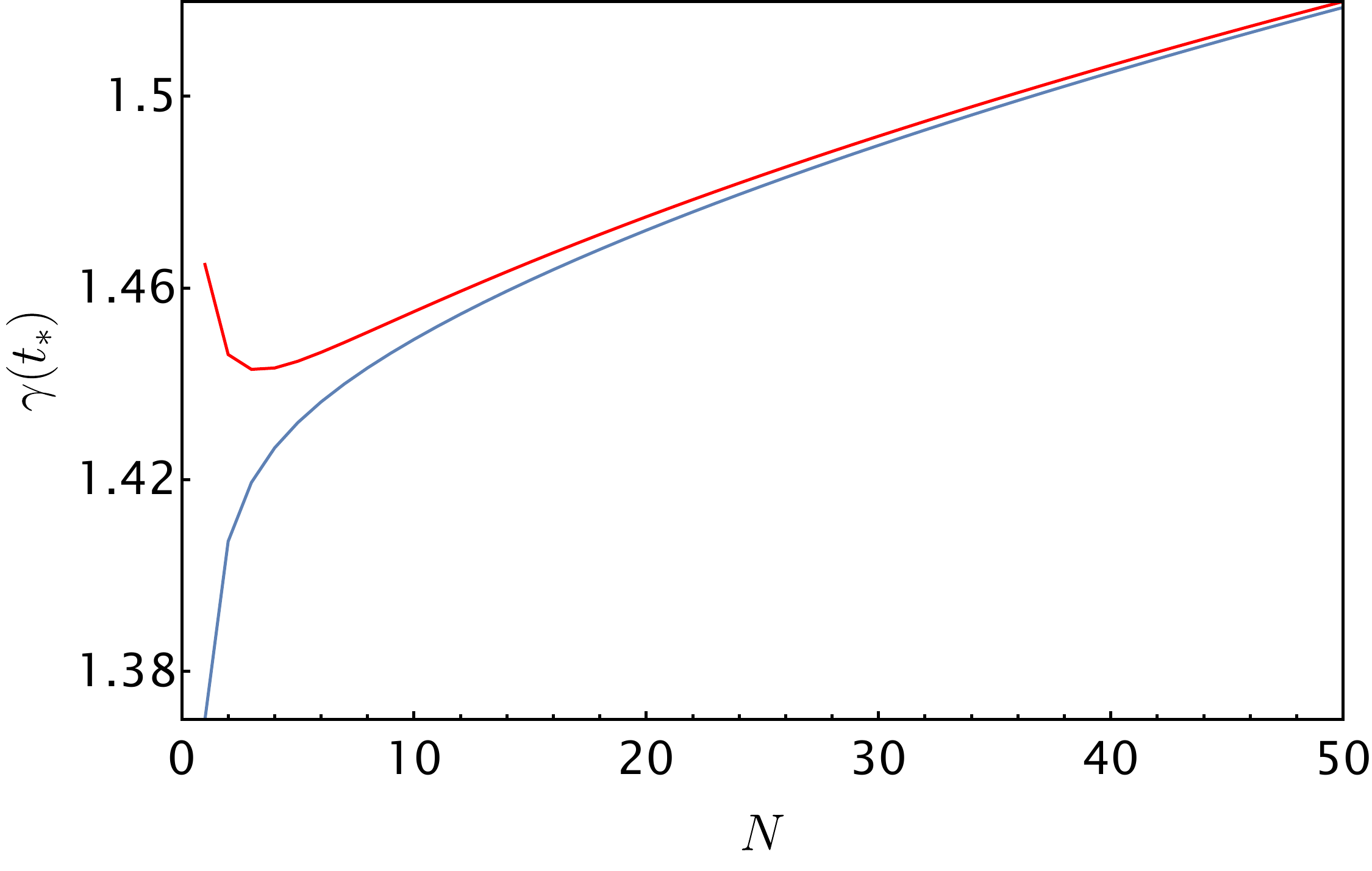}}
\caption{$N$-scaling of the characteristic $\gamma(t_{*})$ for equipartition of energy density at early time. The blue (red) line corresponds to the solution to Eq.~(\ref{eq:GammaNScalingEq}) on the visible (invisible) side of the horizon from the UV observer's perspective. Note that this plot is only valid at whole number values of $N$.}
\label{fig:gammaNscaling}
\end{figure}

Using this estimate, we can associate values of $N$ with particular brane trajectories given the assumption of equipartition at the time where our effective description begins to be valid, $t_{*}$. It is important to note that the metric factor coefficient on the LHS of Eq.~(\ref{eq:GammaNScalingEq}) is a function of $R$ and $t_{*}$, and therefore contributes an additional $N$-scaling to the characteristic value of $\gamma(t_{*})$. There are two solutions for $R(t_{*})$ that satisfy Eq.~(\ref{eq:GammaNScalingEq}), one on either side of the horizon. In Figure~\ref{fig:gammaNscaling}, we show (numerically) how the characteristic value of $\gamma(t_{*})$ scales with $N$ for these two solutions. There is a difference in behavior between the two solutions for small $N$, but as $N$ becomes large, this difference becomes small. 

It is interesting to see that even for larger values of $N$ (corresponding to the 5D gravity theory being more perturbative), the brane remains non-relativistic at early time $t_{*}\sim\Lambda^{-1}$.  As mentioned in Section~\ref{sec:UnstbCosmo}, slower branes enter the horizon at earlier times. In the case of positive vacuum energy, $\delta_0$, and negative dilaton quartic, $\delta_1$, only the highly relativistic trajectories find their way into the metastable attractor where the dilaton vev is driven towards zero. This means that if the energy density is partitioned more or less equally between the dilaton and the other degrees of freedom, we can expect the IR brane to cross the horizon relatively early, avoiding getting stuck in the metastable, inflating configuration, while maintaining perturbative control over the 5D gravity theory. This will be quite relevant for our discussion of the cosmology of the stabilized dilaton, which we address in Section~\ref{sec:StabCosmo}.

%The most interesting feature is the scaling of the relativistic energy at early times with $N$:  the larger the number of colors in the CFT, the lesser the boost factor of the brane at $t_{*}\sim\Lambda^{-1}$.  As mentioned in Section~\ref{sec:UnstbCosmo}, a slower brane enters the horizon at earlier times.  In the case of positive vacuum energy, $\delta_0$, and negative dilaton quartic, $\delta_1$, only very relativistic trajectories find their way into the metastable attractor where the dilaton vev is driven towards zero.  We can interpret our result in this section as predicting that larger $N$ theories will tend to avoid being trapped in the false vacuum.  This will be quite relevant for our discussion of the cosmology of the stabilized dilaton, which we address in Section~\ref{sec:StabCosmo}.

We emphasize again that this naive dimensional analysis estimate comes with the caveat that we have presumed equipartition around the time where our description becomes valid.  We recognize that other UV completions that this cosmology could be matched onto may come with different estimates and $N$ scaling.

%%%%%%%%%%%%%%%%%%%%%%%%%%%%%%%%%%%%%%%%%%%%%%%%%%%%%%%%%%%%%%%%%%%%%%%%%%%%%%%%%%%%
%%%%%%%%%%%%%%%%%%%%%%%%%%%%%%%%%%%%%%%%%%%%%%%%%%%%%%%%%%%%%%%%%%%%%%%%%%%%%%%%%%%%

\section{Stationary de Sitter Solutions}
\label{sec:solutions}
Before we include effects of a Goldberger-Wise (GW) stabilizing scalar field for the full cosmology, we start by mapping out the stationary points of the dilaton potential.  This corresponds to mapping out the effective dynamical mistunes through a large range of $f$, all the way to $f\rightarrow 0$. In this case of stationary points, there is no time-dependence of the scalar field. We are interested not only in the global minimum of the dilaton potential, but also its other stationary points.  

If the true vacuum of the potential has vanishing vacuum energy, the low energy effective theory is a flat 4D spacetime.  At any other stationary points, the vacuum energy should be positive, and we expect to find solutions with 4D de Sitter slices.  The Hubble constant in these inflationary solutions contributes to the shape of the 5D geometry, and feeds into the effective potential for the dilaton.  This effect gives rise to novel features that are not apparent in the static description of the 4D effective potential~\cite{Bellazzini:2013fga}.

To find such solutions, we consider 5D actions of a stabilizing scalar field coupled to Einstein gravity.  
Our ansatz for the 5D metric is
\beq\label{eq:GCoordsAnsatz}
ds^2 = e^{-2ky} \left( dt^2 - e^{2 H  t} dx^2 \right) - k^2 dy^2/G^2(y),
\eeq
where $H$ is the inflationary Hubble rate, and $G$ is a function encoding deviations of the geometry from pure AdS space.  The limit of pure AdS with curvature $k$ is obtained by taking $G = k$, with $H=0$.

We work with scenarios where the geometry is truncated either by two branes (a UV brane at $y_0=0$, and an IR brane at $y_1> 0$), or by a single UV brane at $y_0=0$ and a horizon at finite proper distance from it.  The action is:
\beq
S= \int d^5x \sqrt{g} \left[ \frac{1}{2} (\partial_M \phi)^2- V(\phi) - \frac{1}{2 \kappa^2} \mathcal{R} \right] + S_\text{branes},
\eeq
where $\mathcal{R}$ is the Ricci scalar. We consider simple brane actions of the form
\beq
S_{0,1}=-\int d^4x \sqrt{g_{0,1}} \lambda_{0,1} (\phi),
\eeq
where the IR brane action is not relevant if the geometry is truncated by a horizon.

We now look for solutions to the classical Einstein equations and the scalar equation of motion.  The $yy$ Einstein equation yields an expression for $G$:
\beq\label{eq:GAllOrders}
G^2 = \frac{-\frac{\kappa^2}{6} V(\phi) + e^{2ky} H^2}{1- \frac{\kappa^2}{12} \phi'^2 },
\eeq
and the scalar equation of motion is
\beq
\phi'' =  \left( 4 -\frac{G'}{G} \right) \phi' + \frac{1}{G^2} \frac{\partial V}{\partial \phi}.
\eeq
Plugging in for $G$ gives a non-linear equation which has a somewhat simple form when written using a rescaled Hubble constant $\tilde{H}^2(\phi) = -\frac{6}{\kappa^2} \frac{H^2}{V(\phi)}$:
\beq
\phi'' = 4 \left[ \left(1+ \frac{3}{4} e^{2ky} \tilde{H}^2 \right) \phi' - \frac{3}{2\kappa^2} \frac{\partial \log V}{\partial \phi} \right] \frac{1-\frac{\kappa^2}{12} \phi'^2}{1+e^{2ky}\tilde{H}^2}.
\eeq
In this work, we will concern ourselves primarily with the small backreaction limit of this equation, $\kappa^2 \rightarrow 0$, although in principle this equation could be numerically solved to take into account higher order corrections.

Variation of the action also results in boundary conditions for the scalar field:
\beq
\phi' |_{0,1} = \pm \frac{1}{2 G_{0,1}} \frac{\partial \lambda_{0,1}}{\partial \phi}.
\eeq
Similarly, singular terms in the Einstein equations result in metric junction conditions at the branes:
\beq
G_{0,1} = \pm \frac{\kappa^2}{6} \lambda_{0,1} ( \phi_{0,1} ).
\eeq
Our goal is to find new solutions to this set of equations for input parameters that lead to a stabilized RS geometry.

The scalar equation of motion is non-linear after substituting in the expression for $G$.  However, when the scalar field value is small, we can approximate the potential $V(\phi) \approx - \frac{6 k^2}{\kappa^2}+\frac{1}{2}m^2\phi^2$ and neglect terms of order $\kappa^2 \phi'^2$.\footnote{We hereafter work in units $k = 1$.} With this, the geometry function $G$ has a simple expression:
\beq
G = \sqrt{1+e^{2y} H^2}.
\label{eq:Ginf}
\eeq
We note that the resulting metric has a horizon at $y\rightarrow \infty$ if the geometry is not cut off by an IR brane.  This horizon lives at finite distance from the UV brane:
\beq
L = \int \frac{dy}{G(y)} \approx \log \frac{2}{H}- y_0,
\eeq
where we have assumed $H e^{y_0} \ll 1$.  

In this small-backreaction approximation, and with a simple quadratic potential with mass $m^2$, the scalar equation is linear. It is solved by combinations of the following two solutions:
\beq\label{eq:GScalarSols}
\phi_\pm = e^{(2 \pm \nu)y}  \frac{(G \pm \nu)^{~~~} }{(G+1)^{\pm \nu}},
\eeq
where $G$ is the pure inflationary metric function in Eq.~(\ref{eq:Ginf}), and $\nu = \sqrt{4+m^2}$.

\begin{figure}[t!]
\center{
\includegraphics[width=0.5\textwidth]{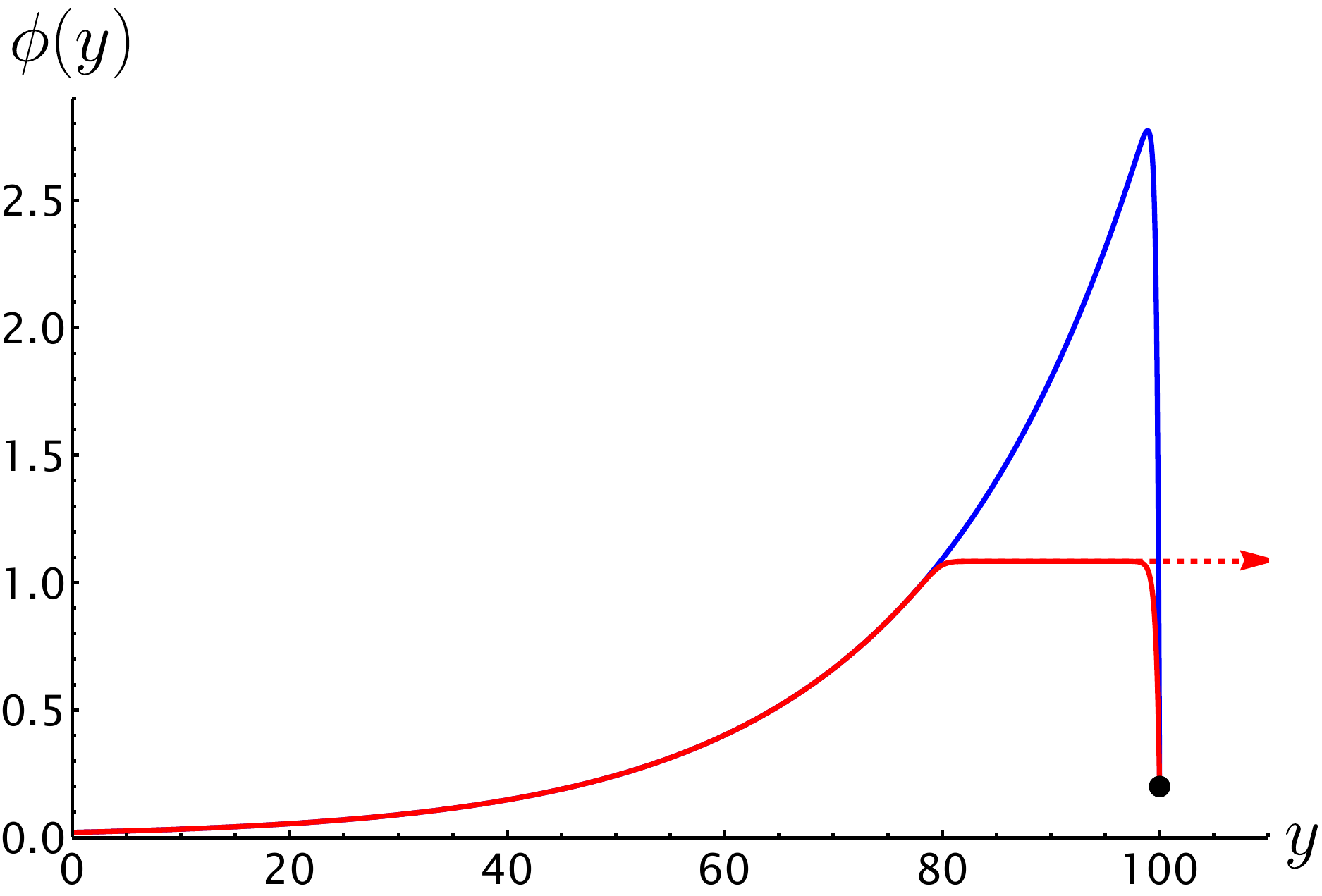}}
\caption{An example of the scalar field evolution for $y_1 = 100$, $v_0 = 1/50$, $v_1 = 1/5$, and $\epsilon = 1/20$.  The blue curve is for $H = 0$, while the red curve exhibits the solution when $\log \mu_0/H \approx 80$.  The dashed arrow indicates the field value as $y\rightarrow \infty$, assuming there is no IR brane, and the solution extends to the horizon.}
\label{fig:plotphi}
\end{figure}

It is good to take a moment to discuss the behavior and interpretation of this solution.  With some example stiff wall boundary conditions, we can compare this solution to one with $H = 0$, which we do in Figure~\ref{fig:plotphi}.  Whereas the $H= 0$ solution allows unlimited exponential growth in the scalar field background, a finite value for $H$ curtails the growth, flattening the solution when $y-y_0 \gtrsim \log \mu_0/H$.  

In regions of $y_1$ that are far from the GW minimum, the flattening due to Hubble expansion can prevent the field growth from creating a bulk region of uncontrolled backreaction.  This allows us to extend solutions all the way to $f\rightarrow 0$, corresponding to the IR brane hiding behind the horizon, and with unbroken conformal symmetry.

%This flattening is consistent with a CFT dual interpretation where the vev breaking the CFT spontaneously is vanishing, and where the RG evolution of the coupling constant is halted when the RG scale passes the threshold associated with the finite size of the universe.  

%Importantly, the IR brane is very close to the horizon when $y_1 -y_0 \gtrsim \log \mu_0/H$ (in proper distance).  As $y_1$ progresses beyond that critical value, it essentially describes the process of the IR brane being swallowed by the horizon, and the dual to the IR brane, the scale of spontaneous conformal symmetry breaking, $f$, does not change significantly.

We separate our search for solutions into two-brane configurations, and solutions where there is only a UV-brane and a horizon terminates the geometry.  

\subsection{Solutions with UV and IR brane}
\label{subsec:2brane}

Our first task is to generate parameters that give a stabilized RS geometry with vanishing 4D effective cosmological constant.  For this, we take $H=0$, decide on a value of $y_1-y_0$ corresponding to a desired hierarchy, and pick some IR and UV brane potentials.  For the purposes of discussion here, we choose
\beq
\lambda_{0,1}(\phi)  = T_{0,1} + \gamma_{0,1} (\phi^2- v_{0,1}^2)^2,
\eeq
and further consider the limit $\gamma_{0,1} \rightarrow \infty$.  This ``stiff-wall'' limit simplifies the boundary conditions for the scalar field:  $\phi(y_{0,1}) = v_{0,1}$.  $T_1$ is determined by the desired value of $y_1$ for the GW minimum, and $T_0$ is then determined by enforcing consistency with the $H=0$ ansatz.  In the case of small backreaction, the brane tensions will not be very far off from the doubly tuned RS values.  We thus take $T_{0,1} = \frac{6}{\kappa^2}\left(\pm 1+ \delta_{0,1}\right)$.

At values of $y_1$ other than the one chosen to be the GW minimum, neither of the metric junction conditions will be satisfied with $H = 0$ -- otherwise we would have missed other degenerate vacua, which is unlikely. An approximate effective potential as a function of $y_1$ is expressible as a pure boundary term~\cite{Bellazzini:2013fga}:
\beq
V_\text{eff} = e^{-4 y_0} \left[ \lambda_0(\phi) - \frac{6}{\kappa^2} G_0\right] + e^{-4 y_1} \left[ \lambda_1(\phi) + \frac{6}{\kappa^2} G_1\right],
\label{eq:effpot}
\eeq
and this can have other maxima and local minima.
However, this does not take into account the fact that a non-vanishing Hubble rate deforms the 5D geometry, and changes the scalar field solution, like in Figure~\ref{fig:plotphi}. This will not typically have a significant effect on such features in this potential, due to the Planck scale suppression of geometric response to energy densities, but it can create features where symmetries broken only by cosmological effects otherwise forbid them. This potential can (and sometimes does) fail in accurately predicting the existence and location of finite vacuum energy extrema.

To find the locations of additional extrema, we include the effects of de Sitter expansion on the solutions, and search again for values of $y_1$ that solve the modified junction conditions.  Consistency requires that the junction conditions in the UV and IR are satisfied independently.  We can re-write the junction conditions as equations that one inverts for the Hubble constant:
\beq
H^2 = e^{-2 y_{0,1}} \left[ \frac{\kappa^4}{36} \lambda^2_{0,1}(\phi_{0,1}) \left( 1- \frac{\kappa^2}{12} (\phi'_{0,1})^2\right) + \frac{\kappa^2}{6} V(\phi_{0,1}) \right].
\label{eq:2BHsols}
\eeq
For the case of stiff wall boundary conditions, the junction conditions simplify, and can conveniently be expressed in terms of the configurations at the GW vacuum solution:
\beq
H^2 = \frac{\kappa^6}{432} T^2_0 e^{-2 y_0} \left[ \left( \bar{\phi}_0' \right)^2 -\left( \phi_0' \right)^2 \right] = \frac{\kappa^6}{432} T^2_1 e^{-2 y_1} \left[ \left( \bar{\phi}_1' \right)^2 -\left( \phi_1' \right)^2 \right],
\eeq
where $\bar{\phi}_i'$ is the value of $\phi'$ on the UV or IR branes when evaluated at the GW minimum, which we have assumed corresponds to vanishing Hubble constant.  Note that the Hubble parameter and the $y_1$ value both appear in the scalar field profile as well.  These relations are simultaneous non-linear equations that one can solve for $y_1$ and $H$.

The power counting is at first rather surprising.  The 4D Hubble rate scales like $M_5^{-9}$.  However, when taking into account the approximate relation between $T_0$ and the bulk potential, the relation between $\kappa$ and the 4D Planck scale, and the interpretation of the scalar field vev in terms of a coupling and a condensate, $f$, which breaks the conformal symmetry, the above is consistent with the 4D Friedmann equation with vacuum energy given by
\beq
\rho_\Lambda \approx 2 \epsilon v_0 \left( v_1 - v_0 \left( \frac{\mu_0}{f}\right)^\epsilon \right) f^4 \left( \frac{\mu_0}{f}\right)^\epsilon,
\eeq
where $f$ is the value of the condensate at the minimum of the GW potential, and $\epsilon \equiv 2-\nu$.  This agrees precisely with the values estimated using the effective potential in Eq.~(\ref{eq:effpot}).

Solutions to Eq.~(\ref{eq:2BHsols}) will correspond to the extrema of the dilaton effective potential that take into account the effects of backreaction of the 4D dS slices onto the 5D geometry and scalar field profile.  Stability of the solutions can be determined by examining the spectrum of states, particularly the holographic dilaton, at these extrema.  We perform this analysis for an example set of parameters in Section~\ref{sec:radionspec}. The GW minimum has a positive mass for the holographic dilaton, indicating its stability. We also find secondary solutions with a larger brane separation than the one at the GW minimum, $H> 0$, and a tachyonic holographic dilaton (unstable).   Finally, there are inflating solutions with no IR brane, corresponding to the case of unbroken conformal symmetry.  The spectrum for the holographic dilaton in this case is a gapped continuum.

The existence of unstable stationary points in the dilaton effective potential is analogous to what we observed in Section~\ref{sec:UnstbCosmo}.  In the situation with positive cosmological constant term, $\delta_0$, and negative quartic $\delta_1$, (second panels in Figs.~\ref{fig:branches} and ~\ref{fig:ExTrajPlots}), the $f=0$ solution is a metastable attractor at late times, when the radiation has dissipated.  In the case of the stabilized dilaton, the potential away from the minimum can be characterized by \emph{effective} mistunes that, for small $f$, have the same pattern of signs.

%In the late-time limit of the unstabilized cosmology for the positive vacuum energy, and negative dilaton quartic scenario, the bottom $\beta=0$ line in the second panel of  Figure~\ref{fig:ExTrajPlots}, that trajectories have to cross to avoid collapsing back to the UV brane, can be associated with the unstable 2-brane stationary point we have identified.

\subsection{Solutions with UV brane and IR horizon}
\label{subsec:braneHor}

Not all geometries involve an IR brane in the physically accessible space (from the perspective of a UV brane localized observer).  The breaking of the AdS isometries with 4D dS slices leads to a finite size bulk terminated by a horizon, and there may be solutions to the scalar-Einstein equations which hide the IR brane beyond the horizon.  

In the coordinates we have chosen, the horizon is at $y\rightarrow \infty$, and provides its own scalar boundary condition.  The asymptotic behavior of the $\phi_\pm$ solutions at large $y$ is $\phi_\pm \propto e^{3y}$ and such scaling would lead to a divergent action.  We conclude that when there is no IR brane visible to a UV brane observer, correct near-horizon behavior enforces a condition that the background solution for the scalar field is
\beq\label{eq:ScalarHorSols}
\phi = c_- \left[ e^{(2-\nu )y} \frac{(G - \nu)^{~~~} }{(G+1)^{- \nu}} - H^{2  \nu} e^{(2+\nu )y} \frac{(G + \nu)^{~~~} }{(G+1)^{+\nu}} \right].
\eeq
This scalar profile is in fact very simple.  It interpolates smoothly between two regimes of evolution.  For $y < \log \mu_0/H$, there is the usual slowly growing exponential behavior dual to near marginal coupling constant evolution.  For $y> \log \mu_0/H$, the vev is simply a constant:
\beq
\phi \approx \left\{ \begin{array}{ll}
v_0 e^{\epsilon y} & y < \log \mu_0/H \nonumber \\
v_0 (\mu_0/H)^{\epsilon} & y > \log \mu_0/H. \end{array}\right.
\eeq

There is no longer a metric junction condition since the geometry is smoothly shut off, so in this case, only the UV brane scalar boundary condition and metric junction condition must be satisfied.  There is no $y_1$ dependence, as the IR cutoff is now given by the Hubble scale:
\beq
H^2 = \frac{\kappa^6}{432} T^2_0 e^{-2 y_0} \left[ \left( \bar{\phi}_0' \right)^2 -\left( \phi_0' \right)^2 \right].
\eeq

In the context of brane models, this treatment is exhaustive, up to the variety of bulk and brane scalar potentials.  All relations (with the exception of the form of the scalar field solution) are valid at all orders in gravitational backreaction, however such equations can't be fully trusted since higher dimensional operators that modify the gravity theory at high energies will begin to play an important role when the scalar field profile significantly deforms the geometry from AdS.

We do find a solution with no IR brane, with $H>0$. This configuration is meta-stable, as indicated by a positive mass-gap in the spectrum of the holographic dilaton. In the late-time limit of the unstabilized cosmology for the positive vacuum energy, and negative dilaton quartic scenario, we can associate the top $\beta=0$ line close to the horizon in the top-right panel of Figure~\ref{fig:branches} with this meta-stable stationary point we have identified.

\begin{figure}[t!]
\center{
\includegraphics[width=0.5\textwidth]{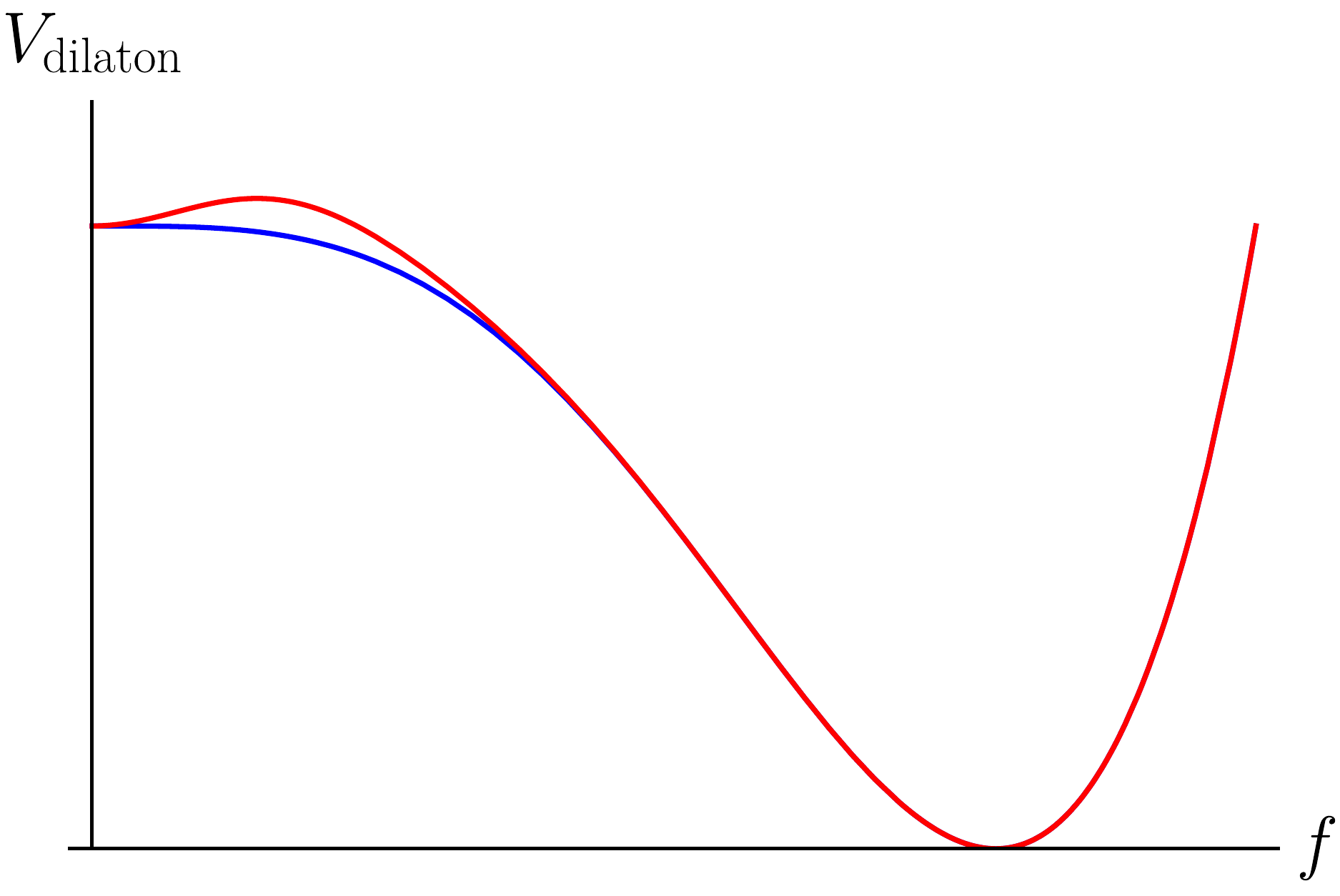}}
\caption{Schematic comparison of the dilaton potential including backreaction from Hubble expansion (red line), and without (blue).}
\label{fig:cartoonPlot}
\end{figure}

To summarize, using a de Sitter ansatz, taking into account the Hubble backreaction, there is one additional solution with an IR brane that is unstable, and one meta-stable solution without an IR brane, where the bulk geometry is shut off by a Rindler-type horizon, with no spontaneous breaking of conformal invariance ($f=0$ configuration). See Figure~\ref{fig:cartoonPlot} for a schematic plot showing all the stationary points in the dilaton potential, upon inclusion of Hubble backreaction.  In Section~\ref{sec:radionspec}, we show the calculation to establish the stability of these stationary points for an example set of parameters.

%%%%%%%%%%%%%%%%%%%%%%%%%%%%%%%%%%%%%%%%%%%%%%%%%%%%%%%%%%%%%%%%%%%%%%%%%%%%%%%%%%%%
%%%%%%%%%%%%%%%%%%%%%%%%%%%%%%%%%%%%%%%%%%%%%%%%%%%%%%%%%%%%%%%%%%%%%%%%%%%%%%%%%%%%

\section{The Spectrum of the Holographic Dilaton}
\label{sec:radionspec}

\begin{figure}
\centering
	\subfloat{\includegraphics[width=0.49\textwidth,keepaspectratio]{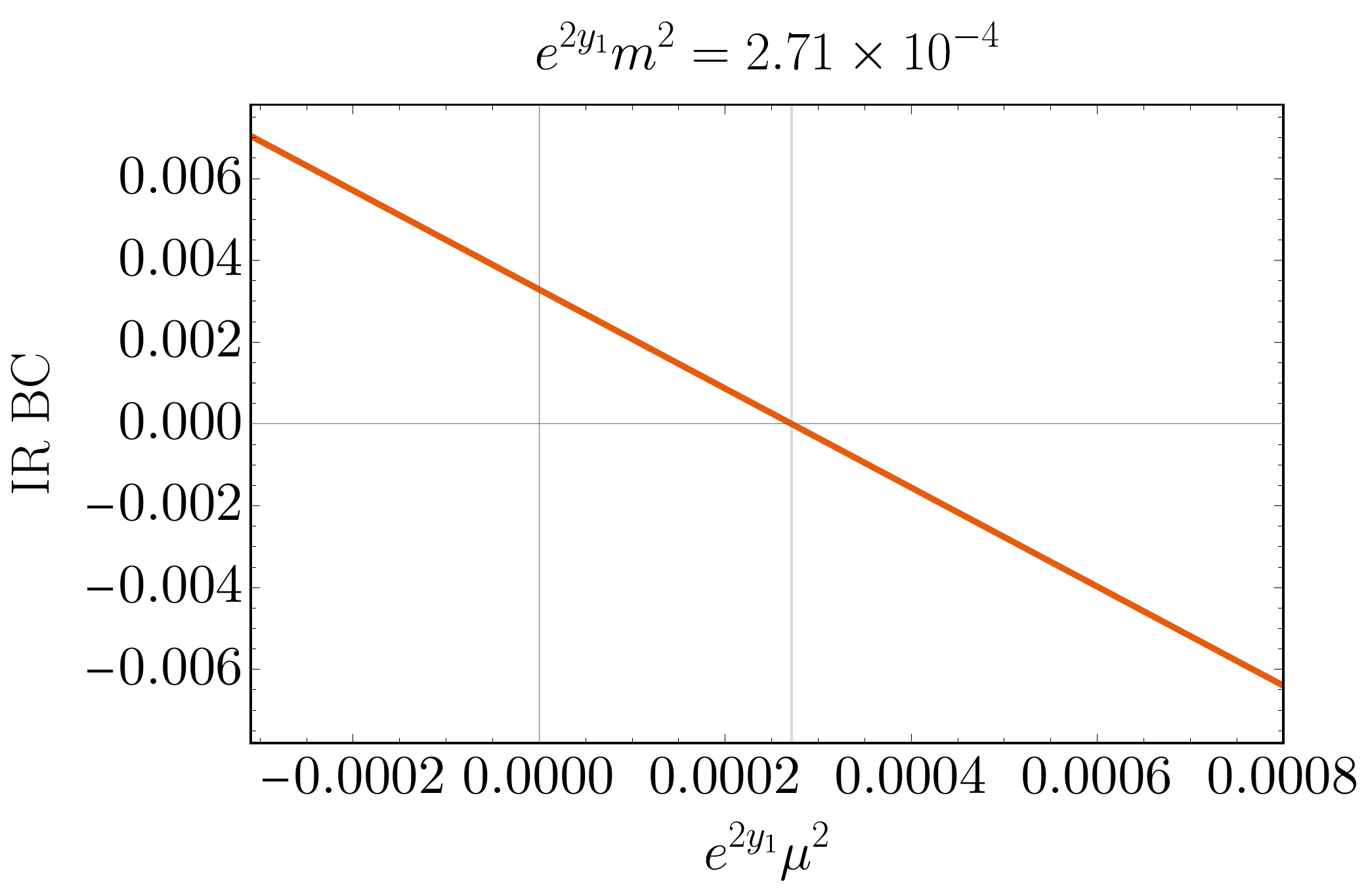}}\hfill%
	\subfloat{\includegraphics[width=0.49\textwidth,keepaspectratio]{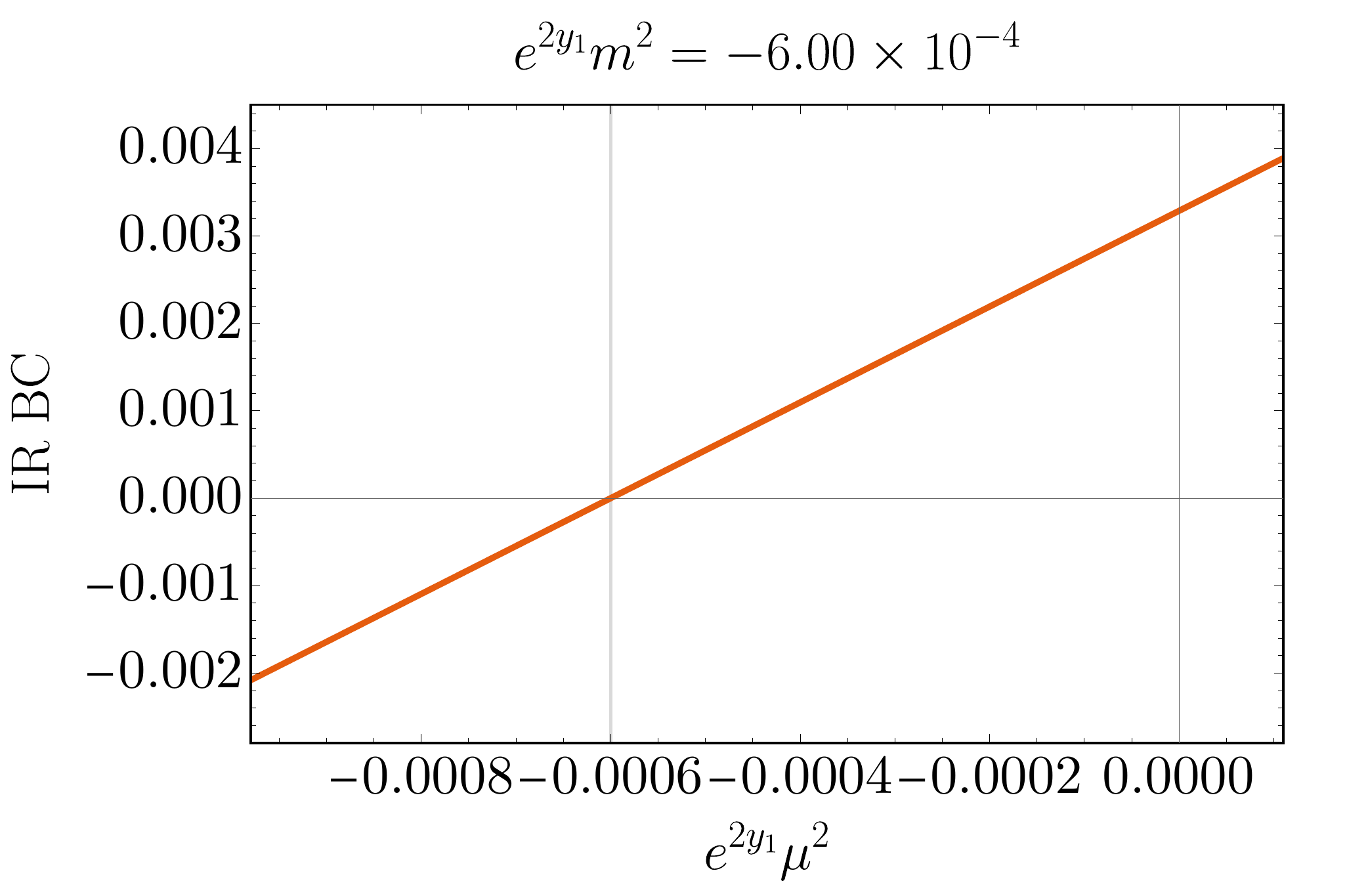}}
\caption{Behavior of the IR boundary condition for the fluctuation as a function of the IR scale. The mass eigenvalue coincides with the location at which the boundary condition is met, i.e.\ with the scale at which the curve displayed crosses zero. The left figure corresponds to the stable GW minimum, the right one with the unstable stationary point. As expected, the mass eigenvalue at the minimum (maximum) is positive (negative). The parameters of the model have been chosen to generate a realistic hierarchy between the Planck (UV) and TeV (IR) scales.}
\label{fig:radion}
\end{figure}

In order to study the spectrum of fluctuations around the background stationary solutions we described in the previous section, we follow the approach of Csáki, Graesser and Kribs~\cite{Csaki:2000zn} and consider the equation for the scalar fluctuations, $F$ (the holographic dilaton) about the background ansatz in Eq.~(\ref{eq:GCoordsAnsatz}):
\beq
\Box_\text{dS} F = G^2 e^{-2y} \left[ F'' -  \left(2+\frac{G'}{G} + 2 \frac{\phi''}{\phi'} \right) F'+ \left( 4 \frac{\phi''}{\phi'} +6H^2\right) F \right],
\eeq
where $\Box_\text{dS}$ is the box obtained by contracting indices with the 4D de Sitter metric. This matches the result of~\cite{Csaki:2000zn}, except of course for the additional term proportional to $H^2$. For the stiff wall boundary conditions, the holographic dilaton fluctuation obeys $F'_{0,1} = 2 F_{0,1}$ on the UV and IR branes.  Using the all-orders expression for $G$ in Eq.~(\ref{eq:GAllOrders}) and the small backreaction solution for the background Eq.~(\ref{eq:GScalarSols}) with stiff-wall boundary conditions, $\phi$, we find, as usual, the massive holographic dilaton solution at the GW minimum.  At the unstable 2-brane stationary point of the dilaton potential, the spectrum is shifted and the holographic dilaton is tachyonic as expected. We demonstrate these points by numerically solving the fluctuation equation and studying the spectrum with a set of parameters that generate a realistic hierarchy between the Planck and electroweak scale. The results are displayed in Figure~\ref{fig:radion}.

To determine the spectrum when there is no IR brane, it is most convenient to work in conformally flat $z$-coordinates where $\exp{y} = \exp(A(z))$, and $dy/G = \exp(-A(z)) dz$.  In this case, the equation of motion for the holographic dilaton becomes
\beq
\Box_\text{dS} F = F'' - F' \left( 3 A' - 2 \frac{\phi''}{\phi'} \right)+ F \left( 4 A' \frac{\phi''}{\phi'} -A'' + 6H^2\right).
\eeq

It is instructive to consider first the case with no scalar field.  In this case, the linearized $\mu5$ Einstein equation can be integrated, yielding $F(z) = \exp(2 A(z))$.  However, this mode is not normalizable in the absence of an IR brane that cuts off the geometry before the horizon.  

A rescaled holographic dilaton field, $\exp(3/2 A)\phi'\tilde{F} =F$ puts the equation of motion into Schrödinger form:
\beq
-\widetilde{F}'' + \left( \frac{9}{4} A'^2 -6H^2 + \frac{5}{2} A'' - A' \frac{\phi''}{\phi'} + 2 \left( \frac{\phi''}{\phi'} \right)^2- \frac{\phi'''}{\phi'} \right) \widetilde{F} = m^2 \widetilde{F}.
\eeq

Asymptotically, the $\phi$ equation of motion can be solved as a series in $e^{-2A}$.  This is due to the fact that $A'' \rightarrow 0$.  Taking $\phi \approx \phi_H + \xi e^{-2A}$ ($\phi_H$ is the background value of the scalar field in Eq.~(\ref{eq:ScalarHorSols}) evaluated at the horizon), we find $\phi''/\phi' \approx -2 A'$, and $\phi'''/\phi' \approx 4 A'^2$.  Plugging in, we find that the asymptotic Schr\"odinger equation (following, for example, ~\cite{Falkowski:2008yr}), in the $y \rightarrow \infty$ limit, is
\beq
-F'' + \frac{9}{4} H^2 F= m^2 F.
\eeq
This leads to a ``gapped continuum'', beginning at scale $\mu = 3/2 H$~\cite{Kumar:2018jxz,Gabadadze:2021dnk}.  The gap is robust under different choices of the GW scalar potential.  The breaking of conformal invariance by the Hubble expansion determines the gap for the spectral density, even though there are other sources of explicit violation of the dual to conformal invariance that affect the high momentum correlation functions.

The key result of the previous two sections is that it is possible in some cases to completely characterize all stationary points of the effective dilaton potential while remaining in the regime of small back-reaction.  These stationary points are characterized by effective mistunes of the brane tensions which are functions of the configuration of the bulk scalar field, and correspond to the dS expansion rate at these points.  In the next section, we show how to utilize these effective mistunes to approximate the dilaton cosmology past the big bang, and to track the evolution of the stabilized dilaton into the basin of attraction of its potential.

%%%%%%%%%%%%%%%%%%%%%%%%%%%%%%%%%%%%%%%%%%%%%%%%%%%%%%%%%%%%%%%%%%%%%%%%%%%%%%%%%%%%
%%%%%%%%%%%%%%%%%%%%%%%%%%%%%%%%%%%%%%%%%%%%%%%%%%%%%%%%%%%%%%%%%%%%%%%%%%%%%%%%%%%%

\section{Stabilized Cosmology}
\label{sec:StabCosmo}
Now that we have identified and characterized the stabilities of the late-time limits of the stationary points in the unstabilized cosmology, we can turn our attention to studying the cosmology in the presence of a Goldberger-Wise stabilizing scalar. For a complete treatment, we would need to solve the coupled system of Einstein equations and the scalar equation with the metric ansatz in Eq.~(\ref{eq:cosmoansatz}), but we instead focus on approximate solutions inspired by slow roll inflationary scenarios.  We begin by writing the equations governing the evolution of the dilaton when the scalar is included.

\subsection{Equations of motion and slow roll parameters}
 We can first employ the $yy$ Einstein equation to show
\beq\label{eq:55eqScalar}
\frac{1}{2} \left[ \frac{1}{n} \dot{\tilde{H}}+ 2\tilde{H}^2 \right] = \frac{1}{2} \frac{a'}{a} \left( \frac{a'}{a} + \frac{n'}{n}\right) +\frac{\kappa^2}{6} V(\phi)  - \frac{\kappa^2}{12} \left[ \phi'^2 + \left( \frac{\dot{\phi}}{n} \right)^2 \right],
\eeq
where $\tilde{H} = \dot{a}/an$.  When this equation is evaluated on the UV brane, the boundary junction conditions allow us to write $a'/a$ and $n'/n$ in terms of brane quantities. The metric junction conditions for the stationary UV brane are
\beq
\left. \frac{a'}{a} \right|_0 = - \frac{\kappa^2}{6} \bar{\rho} ~~~~~~~ \text{and} ~~~~~~~ \left. \frac{n'}{n} \right|_0 =  \frac{\kappa^2}{6} \left( 2\bar{\rho} + 3 \bar{P} \right),
\eeq
where $\bar{\rho}$ and $\bar{P}$ are contributions to the UV brane localized stress energy tensor.   In the case of pure tension, $\bar{\rho} = -\bar{P} = T_0$, and we find $\left. a'/a \right|_0 =\left. n'/n \right|_0 =  - \frac{\kappa^2}{6} T_0$.

Taking $n_0 = 1$ (so $t$ is the proper time for UV brane observers), and enforcing $\dot{\phi} = 0$ on the UV brane for stiff-wall boundary conditions we obtain an equation for the Hubble rate on the UV brane:
\beq
	 \frac{1}{2} \left[ \dot{H}+2 H^2\right]= \frac{\kappa^2}{12} \left[ 1 - \frac{\kappa^2}{12} \left( \phi_0' \right)^2 \right] + \frac{\kappa^2}{6} V (\phi_0) = \frac{\kappa^2}{12}\left[ \left( \bar{\phi}'_0 \right)^2  -\left( \phi'_0\right)^2 \right].
	 \label{eq:hubUV}
\eeq
This equation is exact -- we have not assumed small backreaction due to the effects of the scalar field.  Note that we have dropped the tilde on $H$, since, for our purposes, we define the effective 4D cosmology as the one experienced by UV brane observers:  $\tilde{H}(y=0) \equiv H$.  In the last equality, we have employed the condition that the effective vacuum energy at the Goldberger-Wise potential minimum is vanishing.  Note that this equation is identical to the one obtained in the case of constant mistune $\delta_0$ with the substitution $\delta_0 (2+\delta_0) \rightarrow \frac{\kappa^2}{12}\left[ \left( \bar{\phi}'_0 \right)^2  -\left( \phi'_0\right)^2 \right]$.  Working in the limit of small $\delta_0$ and correspondingly small scalar backreaction we have an effective \emph{dynamical} mistune given by
\beq
\tilde{\delta}_0 \equiv \frac{\kappa^2}{24}\left[ \left( \bar{\phi}'_0 \right)^2  -\left( \phi'_0\right)^2 \right],
\eeq
where $\tilde{\delta}_0$ is now a function of time through its depending on the boundary behavior of the dynamical scalar field.

If the right hand side of Eqn.~(\ref{eq:hubUV}) were constant, the equation could be integrated, giving $H^2 = \frac{4\bar{\lambda}}{\bar{a}^4}+ \frac{\kappa^2}{12}\left[ \left( \bar{\phi}'_0 \right)^2  -\left( \phi'_0\right)^2 \right]$, yielding a solution that is an admixture of radiation and vacuum energy.  We can define a slow-roll parameter for the effective vacuum energy term given by (See Appendix~\ref{sec:AppendixB} for details):
\beq
\epsilon_\text{UV} \equiv  \left| \frac{\dot{\tilde{\delta}}_0}{4 H \tilde{\delta_0}}\right| < 1.
\eeq
When this slow roll condition is satisfied, we can approximate the bulk geometry as the one of constant $\delta_0$ substituting $\tilde{\delta}_0$ in its place.

We now turn to the dynamics of the IR brane. We use a similar set of of relations to obtain an equation determining the evolution of the IR brane position, though these now must account for the non-trivial embedding of the brane (See Appendix~\ref{sec:AppendixA}).
The boundary conditions can again (as we did to obtain Eq.~(\ref{eq:2ndorder})), be plugged into the $yy$ Einstein equation, Eq.~(\ref{eq:55eqScalar}), to get a second order equation for the motion of the brane:
\beq\label{eq:VE2ndorder}
\ddot{R} + \left[ \left( 3 -\frac{6}{\kappa^2 n^2(- T_1)}\frac{\partial V}{\partial R} + \tilde{f}(R,\beta) \right) \frac{\dot{a}}{a} - \frac{\dot{n}}{n} \right] \dot{R} + \frac{\partial V}{\partial R} = 0,
\eeq
where we have pushed all higher order terms in $\beta$ into the function $\tilde{f}$, and the effective potential gradient term is given by
\beq
\frac{\partial V}{\partial R} = \frac{12}{\kappa^2(- T_1)} n^2 \left[  \frac{\kappa^2}{12} \left( \left( \bar{\phi}'_1 \right)^2  -\left( \phi'_1\right)^2\right) -\frac{1}{2} \left(  \frac{1}{n} \dot{\tilde{H}} + 2\tilde{H}^2 \right) \right].
\eeq

Now note that this equation is extremely similar to Eqs.~(\ref{eq:2ndorder}) and (\ref{eq:lindil0}) with the substitution $\delta_1 \rightarrow  \frac{\kappa^2}{24} \left( \left( \phi'_1\right)^2- \left( \bar{\phi}'_1 \right)^2 \right)$.  Also, if the UV slow roll condition is met, the bulk geometry gives 
\beq
n^2 \frac{1}{2} \left(  \frac{1}{n} \dot{\tilde{H}} + 2\tilde{H}^2 \right)  \approx  2 \tilde{\delta}_0 + {\mathcal{O}} (\left(\mu_{\text{IR}}/f\right)^8).
\eeq

This tells us that if this term is relatively constant in time, then the first order equation for the brane evolution, Eq.~(\ref{eq:IRbraneEq}), should be nearly valid if one substitutes in $\delta_1 \rightarrow \tilde{\delta}_1$, with
\beq
\tilde{\delta}_1 \equiv \frac{\kappa^2}{24} \left( \left( \phi'_1\right)^2- \left( \bar{\phi}'_1 \right)^2 \right).
\eeq
We can quantify the validity of this approximation by deriving slow roll conditions for the effective IR brane tension.  The discussion and calculation can be found in Appendix~\ref{sec:AppendixB}, and results in two slow roll parameters corresponding to two epochs.

If we have $\frac{\dot{a}}{a} > \frac{a'}{a} \dot{R}$, then we must satisfy
\beq
\eta_\text{IR} \equiv \left| \frac{n^2 \tilde{\delta}'_1}{3 \left( \dot{a}/a \right)^2} \right| < 1,
\eeq
while in the other regime, when $\frac{\dot{a}}{a} < \frac{a'}{a} \dot{R}$, we instead must satisfy
\beq
\epsilon_\text{IR} \equiv \left| \frac{\tilde{\delta}'_1}{4 \left( \tilde{\delta}_1+ e^{2R} \tilde{\delta}_0 \right) } \right| < 1.
\eeq
These slow roll conditions are valid throughout a significant regime of the majority of trajectories of interest for our study.  
%We indicate in Figure~\ref{fig:RdepRegValid} the region in which all slow roll conditions are satisfied.  
Most of the regions we are most interested in satisfy the slow roll conditions, with the exception being primarily near the minimum of the effective potential where we expect oscillatory behavior of the solutions.  This is sufficient for our purposes, since we are concerned in this work primarily with success or failure of the system to reach the global minimum of the potential.  In the next section, we characterize the brane cosmology for a specific set of parameters, and display the region of validity of the slow roll approximation.

For comparison, in slow roll inflation, the slow-roll conditions are violated at the exit of inflation, where it is presumed the inflaton quickly finds the minimum of its potential, and its oscillations contribute to reheating through decay to light particles.    In our analysis, the portion of the dynamics for trajectories near the minimum of the stabilizing potential is not in the slow roll regime, and we do not expect that using the first order equation with effective constant $\delta_0$ and $\delta_1$ will accurately capture the ``ring-down'' of the dilaton, which may or may not include oscillations, which we expect would decay to SM particles in realistic scenarios.

\subsection{Cosmology of the stabilized Goldberger-Wise model}

To put this on more concrete footing, consider a scalar field with bulk mass term $m^2=\epsilon(\epsilon-4)$, which gives the usual background solution (ignoring the sub-leading effects due to Hubble expansion):
\beq
\phi=\phi_{\epsilon}e^{\epsilon y}+\phi_4e^{(4-\epsilon)y}.
\eeq
If we take stiff-wall boundary conditions $\phi(y=0) = v_0$, and $\phi(y=R(t)) = v_1$, we can use this solution to determine values for the effective mistunes:
\begin{align}
\tilde{\delta}_0(R)&=\frac{\kappa^2}{3} \epsilon v_0 v_1 e^{-(4-\epsilon) \bar{R}} \left[ \left( 1- e^{-(4-\epsilon) (R-\bar{R})} \right) - \frac{v_0}{v_1} e^{\epsilon \bar{R}} \left( 1- e^{-(4-2\epsilon) (R-\bar{R})} \right) \right], \nonumber \\
\tilde{\delta}_1(R)&= \frac{4 \kappa^2}{3} v_0 v_1 e^{\epsilon \bar{R}} \left[\left( 1 - e^{\epsilon (R-\bar{R})} \right) - \frac{1}{2}\frac{v_0}{v_1} e^{\epsilon \bar{R}} \left( 1 - e^{2\epsilon (R-\bar{R})} \right) \right].
\end{align}

\begin{figure}[h!]
\center{
\includegraphics[width=0.7\textwidth]{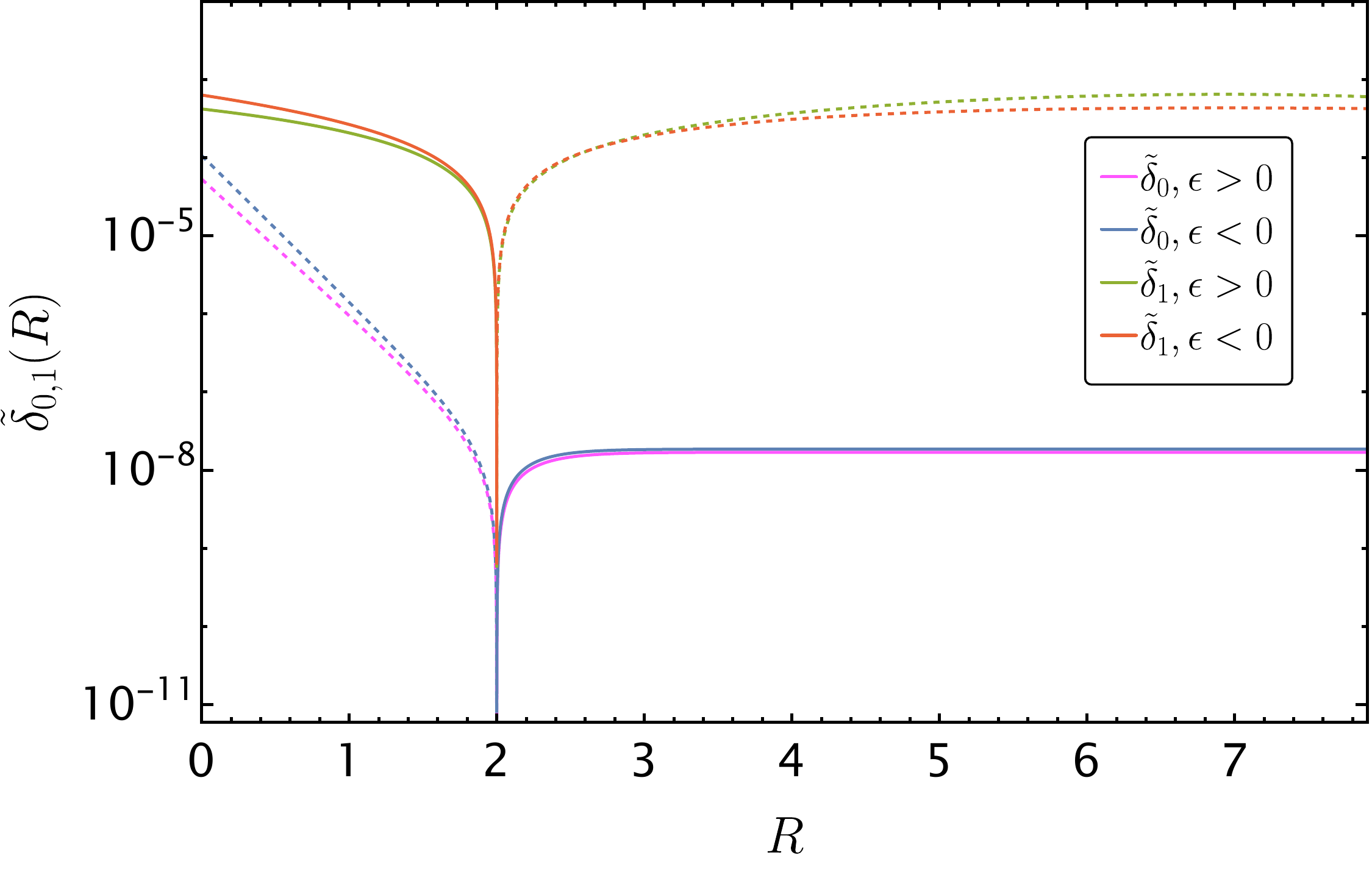}}
\caption{In this plot, we show the values of $|\tilde{\delta}_{0,1}(R)|$ for parameter values of $v_0 =\frac{1}{5\sqrt{3}}$, $v_0=\frac{2}{5\sqrt{3}}$, $|\epsilon|=\frac{1}{10}$, and $\kappa=\frac{1}{2}\sqrt{\frac{3}{2\pi}}$ (corresponding to $N=3$ and $\bar{R} = 2$).   We indicate when the mistunes are negative with dashed lines, and positive with solid lines.  We note that regardless of the sign of $\epsilon$, the effective cosmological constant is positive when $R$ is greater than its value at the stabilized minimum.  The solution with $f\rightarrow 0$  is thus one with an inflating FRW universe.}
\label{fig:d0d1Rdep}
\end{figure}

With these, we can study the stabilized cosmology for both signs of the bulk mass parameter $\epsilon$. In Figure~\ref{fig:d0d1Rdep}, we show how the effective vacuum energy $\tilde{\delta}_{0}$ and the effective dilaton quartic $\tilde{\delta}_{1}$ vary with the position of the IR brane, $R$ for either sign of $\epsilon$. Note that we require the quantity $(v_1-v_0)\epsilon$ to be greater than zero for the dilaton to be non-tachyonic.  Of note is the fact that, independent of the sign of the bulk mass term, the effective dynamical mistunes are of the $-+$ type (panel 3 of Fig.~\ref{fig:branches} when $R < \bar{R}$, and of the $+-$ type (panel 2 of Fig.~\ref{fig:branches}) when $R > \bar{R}$.

\begin{figure}[h!]
\center{
\includegraphics[width=0.49\textwidth]{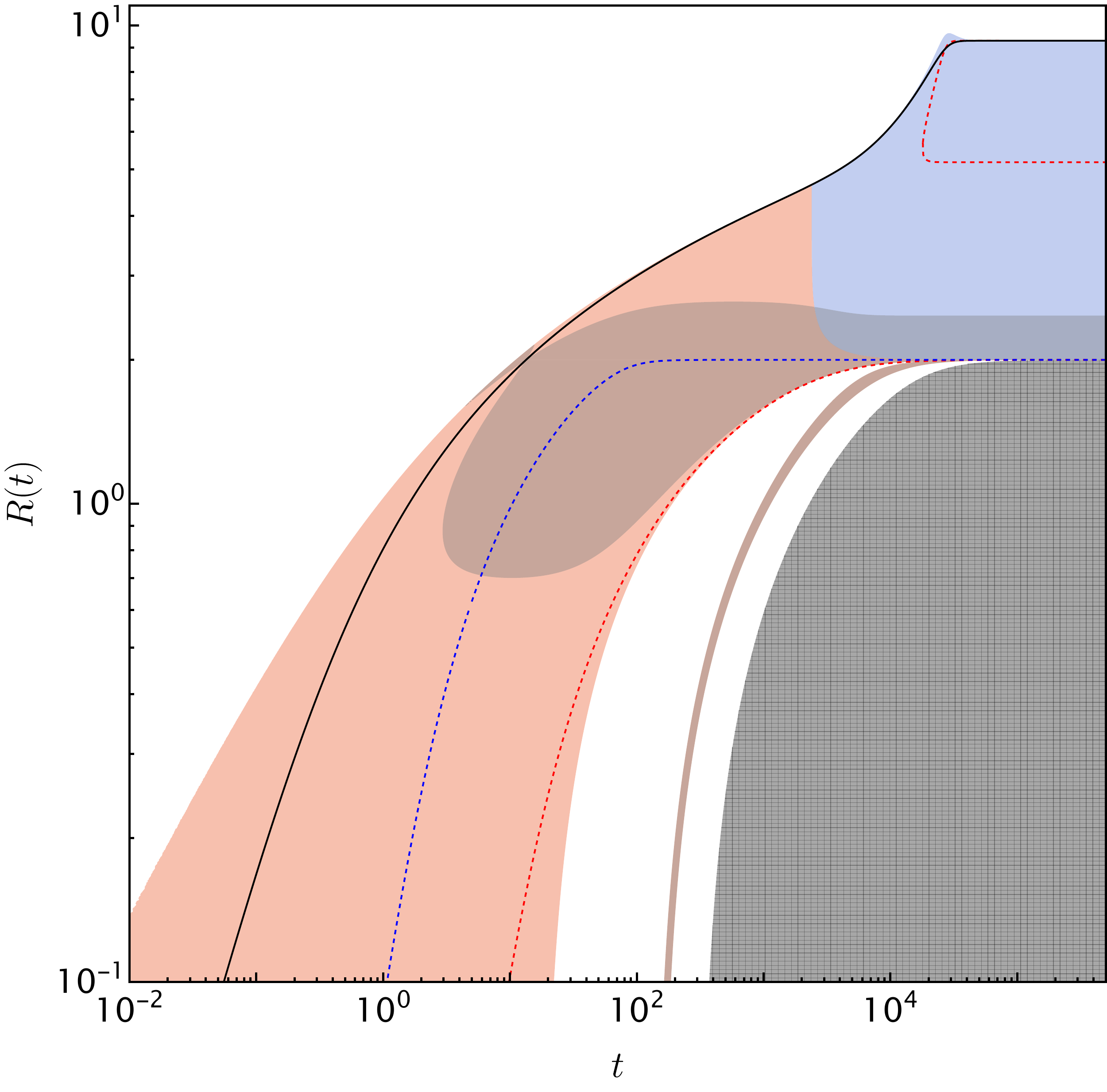}
\includegraphics[width=0.49\textwidth]{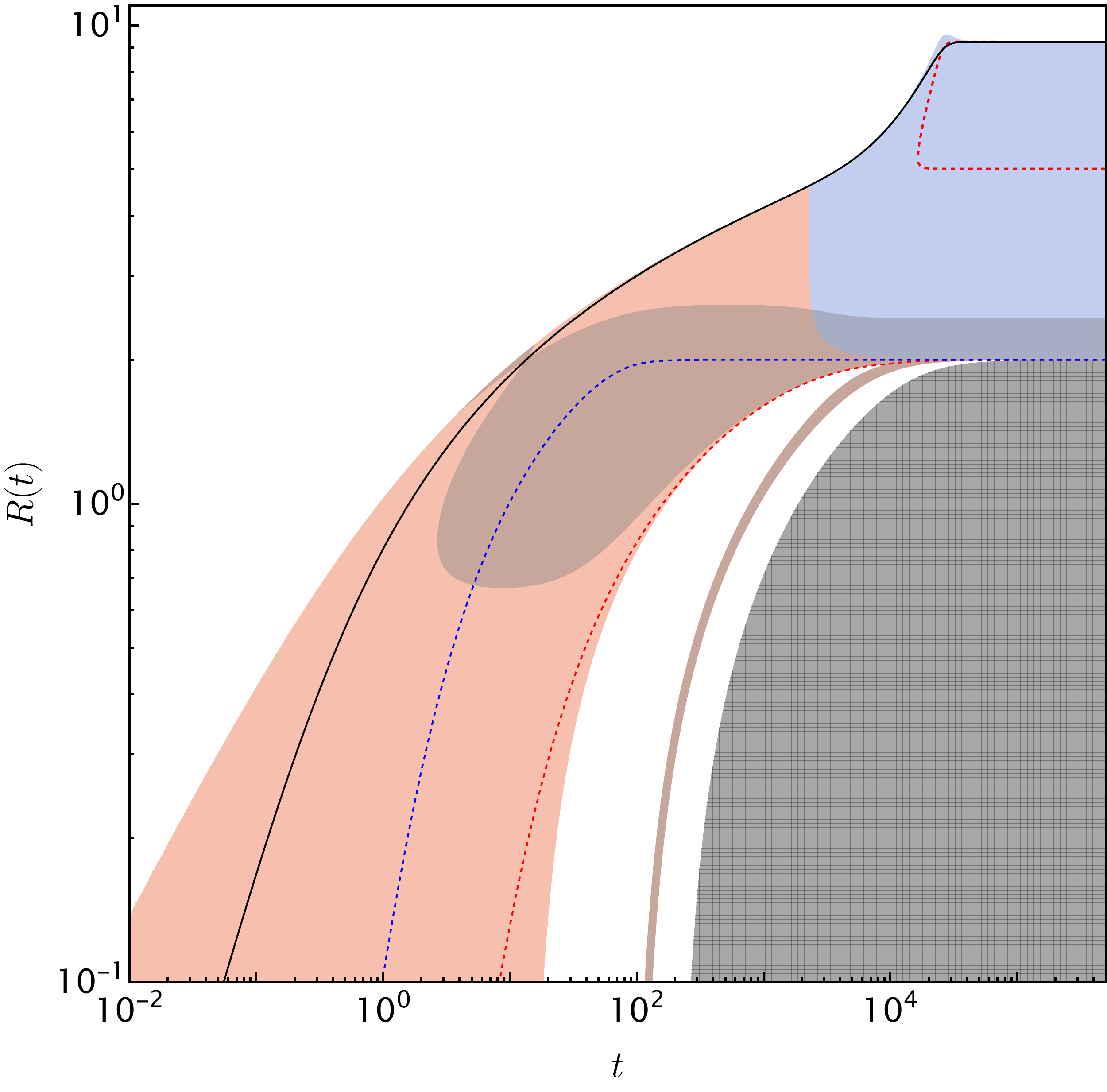}}
\caption{Characterization of the brane cosmology for positive bulk mass (left), and negative bulk mass (right). Red (Blue) regions are dominated by dark radiation (cosmological constant); the black lines are the Rindler horizons; red dotted lines are the turnaround points for trajectories; the blue dotted line is the Goldberger-Wise minimum evolving with time; the plain grey shaded region is where the slow-roll conditions for $\tilde{\delta}_{0,1}$ are not met; the mesh textured grey region are disjoined regions in the bulk space-time, where the brane cannot reach.}
\label{fig:RdepRegValid}
\end{figure}

In Figure~\ref{fig:RdepRegValid}, we characterize all bulk regions, indicating features of interest.  The minimum of the effective potential, obtained by finding the zero of the effective $\partial V/\partial R$ term in Eq.~(\ref{eq:VE2ndorder}), is shown as a dotted blue line. We have shaded the region in which the slow-roll conditions are violated, and we see that this region is concentrated around the minimum of the effective potential.  Red dashed lines correspond to turnaround points for the brane -- where $\dot{R}$ is zero in the slow-roll approximation.  The red region corresponds to the epoch of radiation domination for the UV observer, while the blue region is where the effective cosmological constant term dominates.  There is a region inaccessible to the brane with a great deal of structure caused by the tendency of the universe to crunch when the effective cosmological constant mistune is negative.  We have hidden this underneath the dark gray hatched region in the bottom right of these plots.  Note that in this plot, we have not chosen phenomenological parameters for the GW potential, setting its late-time minimum to be at $\bar{R} = 2$.  Numerical calculation is difficult to stabilize for large hierarchies.  The only difference, however, is for the metastable attractor region in the upper right to be pushed to hierarchically larger times, and to larger values of $y$, while the minimum of the effective potential evolves towards $\bar{R} \approx \log 10^{16}$.

Now, we can return to the question of what ``natural'' trajectories would look like, and what the ultimate fate of the stabilized dilaton will be. If trajectories fall into the meta-stable solution, the cosmology experiences a period of inflation that can only terminate through tunneling to the true vacuum, which is the stabilized Goldberger-Wise configuration.  One must then calculate the bubble nucleation rate and compare with the expansion rate to determine if the transition will actually occur.  This is the usual issue with the conformal phase transition, where the twin requirements of perturbativity of the 5D theory and a rapid transition rate are in tension.  We would like to avoid this problem entirely. If most trajectories in the cosmologies we study end up in the GW minimum directly, the question of transition rate does not come into play.   

\begin{figure}[t!]
\center{
\includegraphics[width=0.49\textwidth]{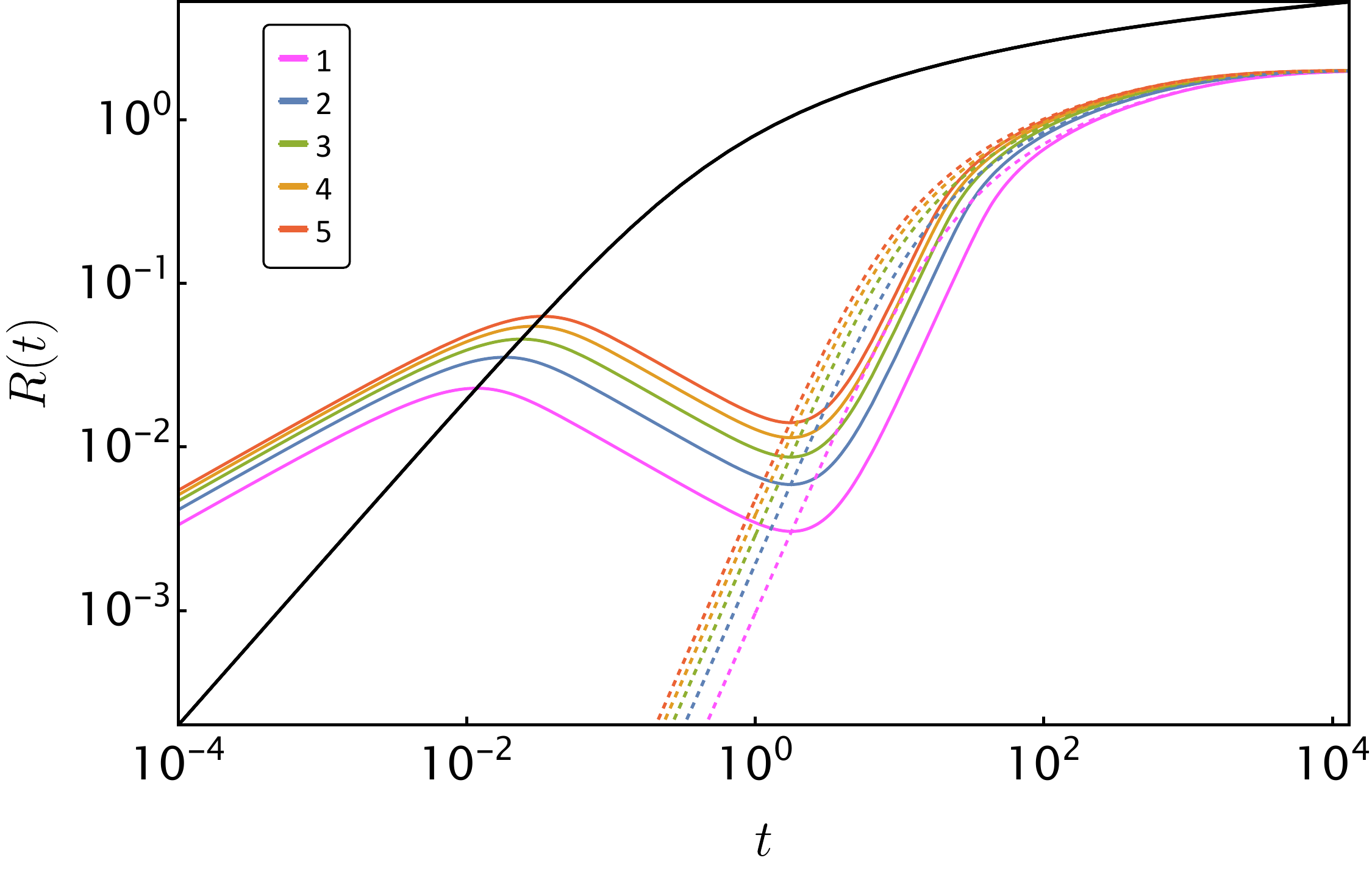}
\includegraphics[width=0.49\textwidth]{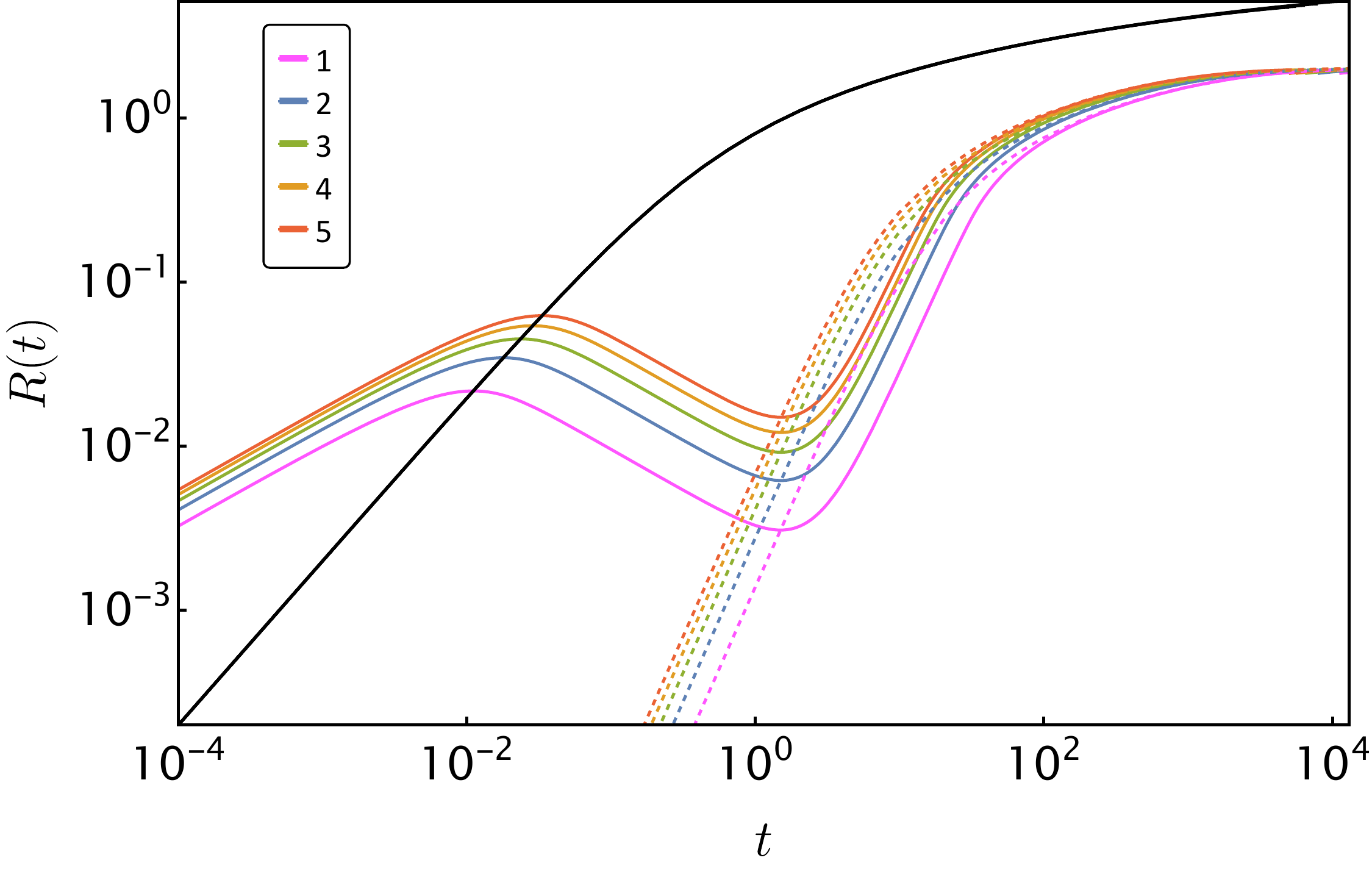}}
\caption{Brane trajectories for $\epsilon>0$ (left), $\epsilon<0$ (right) with equipartition of energies at early times. The dotted lines are the turnaround points that asymptote to the GW minimum at late time, the solid colored lines are the trajectories for a given $N$ (see legend), and the solid black line is the Rindler horizon. The solution to Eq.~(\ref{eq:GammaNScalingEq}) on the visible side of the horizon is chosen for these plots.}
\label{fig:NatTrajPlots}
\end{figure}

In Figure~\ref{fig:NatTrajPlots}, we display trajectories for various choices for model parameters.  We refer to Section~\ref{sec:NatTraj} for our discussion on our choice for initial conditions for the trajectories given a choice of the number of colors, $N$, of the dual CFT.   In this figure, note that $\kappa^2=\frac{N}{8\pi m^3_\text{Pl}}$ has a scaling with $N$, giving slightly different evolutions for $\tilde{\delta}_{0,1}(R)$ and in turn, the corresponding GW minima and turn-around points (shown in plots). We see that most trajectories indeed end up at the GW minimum and completely avoid the meta-stable configuration. To quantify this statement, consider an example set of parameters that resolve the Planck-weak hierarchy- $\bar{R}=\log(10^{16})$, $v_0=1/100$, $v_1=1/2$, $\epsilon=1/10$, $N=5$. If the energy density at early time is partitioned equally between the degrees of freedom, as in Eq.~(\ref{eq:EqPartition}), the brane crosses the horizon relatively early, and becomes non-relativistic at a time associated with the 5D cutoff scale (defined in Eq.~(\ref{eq:5DCutOff})): $\gamma(t_*)\approx 1.43$, and settles to the Goldberger-Wise global minimum of the dilaton potential. On the other hand, for the IR brane to make it to the meta-stable $f=0$ configuration, the brane has to be highly relativistic at early times: $\gamma(t_*)\approx 1.82\times 10^4$, requiring a significant disparity in the partitioning of energies at early time towards the dilaton degree of freedom.

This is a primary conclusion of our work - the cosmological dynamics are such that the IR brane typically enters the horizon during a radiation dominated phase where the effective potential has \emph{not yet formed a metastable minimum at $f \rightarrow 0$.}  The brane thus seeks out the global Goldberger-Wise solution at quite early times, and reaches this configuration well before the period when the universe has cooled enough to reveal the presence of the meta-stable $f=0$ configuration. 

%%%%%%%%%%%%%%%%%%%%%%%%%%%%%%%%%%%%%%%%%%%%%%%%%%%%%%%%%%%%%%%%%%%%%%%%%%%%%%%%%%%%
%%%%%%%%%%%%%%%%%%%%%%%%%%%%%%%%%%%%%%%%%%%%%%%%%%%%%%%%%%%%%%%%%%%%%%%%%%%%%%%%%%%%

\section{Conclusions and Future Work}

We have considered 5D dynamical cosmological solutions of the stabilized holographic dilaton and their role in completion of the conformal phase transition.  This analysis corresponds, via the AdS/CFT dictionary, to a study of out-of-equilibrium dynamics where trajectories of the dilaton do not depend solely on thermodynamic quantities in the early universe, but have sensitivity also to initial conditions.

We find that the cosmologies of the unstabilized dilaton yield insight into the behavior of stabilized models in many regimes of evolution, and so we have thus carefully elaborated on the dynamics of these scenarios, with different relative signs of the UV and IR mistunes.  We have also commented on the AdS/CFT interpretation of features of these solutions, where the UV and IR mistunes correspond, respectively, to a vacuum energy term and a scale invariant quartic for the dilaton.
At early times, all of these cosmologies are dark radiation dominated, with the late time cosmology depending on the sign of the vacuum energy term.  The dynamics of the dilaton are not solely determined by the dilaton quartic, but instead are sensitive also to the vacuum energy through backreaction on the geometry due to the Hubble expansion rate.

The dynamics of the stabilized dilaton are well-approximated by slowly evolving mistunes in many regions of cosmological evolution, and we have provided slow-roll conditions where this approximation is valid. This yields brane-tension mistunes that vary with the location of the IR brane, and which simultaneously vanish at the global minimum of the zero-temperature dilaton effective potential.  

The ultimate fate of the dilaton is thus determined by initial conditions.  At early times, the trajectories of the IR brane universally become highly relativistic, and are obscured from UV brane observers by a Rindler horizon.  The passage of the brane through the horizon corresponds to a phase transition where the dilaton develops a vacuum expectation value which then evolves with time.  We find that many such trajectories move quickly to the global minimum of the effective potential, and it is only trajectories with extraordinarily high initial energy isolated in the dilaton degree of freedom that become trapped in a false vacuum at the origin for the dilaton.

We do not offer a full matching of our study onto other early universe dynamics, such as inflation and reheating.  However, the generic conclusion is that the transition easily completes so long as the partitioning of energy it not very strongly biased to reside in the dilaton degree of freedom.  This is due to the effect of dark radiation on the effective dilaton potential which erases the metastable configuration at early times, clearing the way for the dilaton to move to the global minimum of its potential so long as it is not very highly relativistic at early times.

Absent from our analysis is a study and discussion of the phenomenology, which we leave for future work.  As the transition appears smooth and without violent nucleation and subsequent collisions of true vacuum bubbles, we might guess that the gravity wave signatures associated with a first order transition are absent.  We have not, however, considered the evolution of initial spatial fluctuations of the brane prior to entry through the Rindler horizon.  Such fluctuations are sure to exist, being sourced initially by fluctuations during inflation, and by whatever dynamics precedes the history we describe.  It would be interesting to understand how such fluctuations evolve along the trajectories we describe, and whether they lead to interesting cosmological signatures in the CMB, and in the spectrum of primordial gravitational waves.  We additionally note that our approximation breaks down in a particularly interesting region, where the dilaton moves to the minimum of its potential, and presumably undergoes a dissipation of its energy into light particles.  This process might be trajectory dependent, with two broad classes of initial conditions separating different behavior - those where the trajectories enter during either radiation or vacuum energy domination.

%%%%%%%%%%%%%%%%%%%%%%%%%%%%%%%%%%%%%%%%%%%%%%%%%%%%%%%%%%%%%%%%%%%%%%%%%%%%%%%%%%%%
%%%%%%%%%%%%%%%%%%%%%%%%%%%%%%%%%%%%%%%%%%%%%%%%%%%%%%%%%%%%%%%%%%%%%%%%%%%%%%%%%%%%

\section*{Acknowledgements}
We thank Csaba Cs\'aki for helpful discussions during this work. We especially thank Ameen Ismail for for useful collaboration at early stages of this study.  Jay Hubisz, Bharath Sambasivam, and Gabriele Rigo thank Cornell University for hospitality.  Jay Hubisz and Bharath Sambasivam are supported in part by U.S. Department of Energy (DOE), Office of Science, Office of High Energy Physics, under Award Number DE-SC0009998. Gabriele Rigo acknowledges funding from the European Union’s Horizon 2020 research and innovation programme under the Marie Skłodowska-Curie actions Grant Agreement no 945298-ParisRegionFP. The research activities of Seung J. Lee are supported by the  the National Research Foundation of Korea (NRF) grant funded by the Korea government (MEST) (No. NRF-2021R1A2C1005615), and also by the Samsung Science Technology Foundation under Project Number SSTF-BA2201-06. The research activities of Cem Er\"{o}ncel are supported by 2236 Co-Funded Brain Circulation Scheme2 (CoCirculation2) of The Scientific and Technological Research Council of Turkey T\"{U}B\.{I}TAK (Project No: 121C404). 

\appendix

%%%%%%%%%%%%%%%%%%%%%%%%%%%%%%%%%%%%%%%%%%%%%%%%%%%%%%%%%%%%%%%%%%%%%%%%%%%%%%%%%%%%
%%%%%%%%%%%%%%%%%%%%%%%%%%%%%%%%%%%%%%%%%%%%%%%%%%%%%%%%%%%%%%%%%%%%%%%%%%%%%%%%%%%%

\section{5D Cosmology Solutions}
\label{sec:AppendixA}
In this appendix, we review the solutions to the 5D background cosmology, show limits relevant for various cosmological epochs, and discuss the effects of scalar backreaction. We also derive the IR brane junction conditions. In this appendix, we work in units where the AdS curvature $k=1$.
\subsection{Pure gravity solution}
The $ty$ component of the Einstein equations give an equation that relates the functions $n$ and $a$, giving
\beq
\frac{n}{n_0} = \frac{\dot{a}}{\dot{\bar{a}}},
\eeq
where $\bar{a} = a(y=0,t)$ is the scale factor evaluated at the location of the UV brane.
We make the choice that the cosmology on the UV brane is in proper time coordinates: $n_0 = 1$.
The $tt$ Einstein equations separate out the dependence of the metric function $a$ on $t$ and $y$, yielding
\beq
a^2 = \Lambda_+(t) e^{2y} +  \Lambda_-(t) e^{-2y}-\frac{1}{2} \dot{\bar{a}}^2.
\eeq
 The difference of the $tt$ and $yy$ equations yields a relation between time-dependent pieces which can be integrated to give
\beq
\Lambda_+ = \left[ \left( \frac{\dot{\bar{a}}}{2} \right)^4 + \bar{\lambda} \right] \frac{1}{\Lambda_-},
\eeq
where $\bar{\lambda}$ is an integration constant.  It encodes (from the UV brane perspective) the amount of radiation in the universe, as we will see below.

We might now apply the boundary conditions.  We have a boundary condition relating $a'$ to the UV brane tension:  $-\bar{a}'/\bar{a} = \frac{\kappa^2}{6} T_0$, whose time derivative gives the same relation.

Let us consider a mistune $\delta_0$:  $\frac{\kappa^2}{6} T_0 = 1+\delta_0$.  The boundary condition imposes
\beq
\Lambda_-(t) = \left( \frac{1+ \delta_0}{\delta_0} \right) \left( \frac{\dot{\bar{a}}}{2}  \right)^2 \pm \frac{1}{\delta_0} \sqrt{ \left( \frac{\dot{\bar{a}}}{2}  \right)^4 - \delta_0 ( 2+ \delta_0 ) \bar{\lambda}}.
\eeq
It turns out that on the solution, the argument in the square root is always positive.  Additionally the choice of sign on the two branches is dictated by continuity of the solutions across equality of the two terms in the square root.  

The metric is now completely specified in terms of what is so far an arbitrary function of time, $\bar{a}(t)$, and its derivative.

Now note that there is a consistency relation.  We must have $a^2(y=0,t) = \bar{a}^2$.  The left hand side is purely a function of time derivatives of $\bar{a}$, so this relation gives a differential equation for $\bar{a}$.  The equation can be solved analytically to get $\dot{\bar{a}}$ in terms of $\bar{a}$:
\beq
H^2 = \left( \frac{\dot{\bar{a}}}{\bar{a}} \right)^2 = \frac{4 \bar{\lambda}}{\bar{a}^4} + \delta_0 ( 2 + \delta_0 ).
\eeq
This is the Friedman equation for a universe that includes vacuum energy and radiation.   
    
The full $(t,y)$ dependent solution then stems from the equation for $\bar{a}$:
\beq
\bar{a}^2(t) =\frac{2 \sqrt{\bar{\lambda}}}{\Lambda_4} \sinh ( 2 \Lambda_4 t),
\eeq
where $\Lambda_4 \equiv  \sqrt{\delta_0 (2+ \delta_0)}$.
There is an integration constant that shifts the definition of $t=0$ that has been ignored - the singularity is at $t=0$.  For tiny $\Lambda_4$, or early times this reproduces the usual behavior for radiation domination, $a_0 \propto \sqrt{t}$.  At late times, the usual de Sitter (for positive $\delta_0$) expansion takes over.

  In the case where a positive vacuum energy is dominant at late times, for $\delta_0 >0$, we have:
\beq
y_H \rightarrow \frac{1}{2} \log \left( \frac{2+ \delta_0}{\delta_0} \right).
\eeq
%It is probably possible to show it analytically, but it is anyway clear that there is a Rindler type horizon when you plot solutions for various values of $\delta_0$.  The horizon first proceeds to the ``right," but it eventually asymptotes to a particular value of $y$ that grows as you pick $\delta_0$ smaller and smaller.  This all makes sense - the UV observer on the brane starts off in radiation domination, experiencing acceleration of slower values until the CC takes over, and a period of constant acceleration begins.  At constant acceleration, the Rindler horizon should be at fixed distance from the accelerating observer.
%

It is instructive to consider the two distinct epochs:  radiation domination and vacuum energy domination (for UV brane observers).  In the case of radiation domination, we can consider the limit in which $\delta_0$ is small.  In this case, setting $\bar{\lambda} =1$, the metric functions are
\begin{align}
\bar{a}^2 & = 4 t, \nonumber \\
\Lambda_- &=  \left( \frac{1}{4t} + 4t \right), \nonumber \\
\Lambda_+ &= \frac{1}{4t}, \nonumber \\
a^2(y,t) &=\frac{1}{t} \sinh^2 y + 4 t e^{-2y}, \nonumber \\
n^2(y,t) & = \frac{1}{2 a \dot{\bar{a}}} \left[ 4 e^{-2y} - \frac{1}{t^2} \sinh^2 y \right].
\end{align}  
The $n$ function has a zero whose position changes with time, corresponding to a Rindler type horizon.  In the UV brane radiation dominated era, the location of this horizon is given by
\beq
y_H(t) = \frac{1}{2} \log \left( 1 + 4 t \right).
\eeq

When the cosmology is dominated by positive vacuum energy, the behavior of $a$ is ``4D,'' in that $a^2$ completely separates:
\begin{align}
\bar{a}^2 & = e^{2 \sqrt{\delta_0 (2+\delta_0)} t}, \nonumber \\
a^2(y,t) &=\delta_0 (2+\delta_0) \bar{a}^2\sinh^2 \left[ \frac{1}{2} \log \frac{2+\delta_0}{\delta_0} -y \right], \nonumber \\
n^2(y,t) & = \delta_0 (2+\delta_0) \sinh^2 \left[ \frac{1}{2} \log \frac{2+\delta_0}{\delta_0} -y \right].
\end{align} 
There is a horizon in this case, at $y_H = \frac{1}{2} \log \frac{2+\delta_0}{\delta_0}$.

The full solution interpolates smoothly between radiation domination and domination by the vacuum energy term.

\subsection{Including backreaction}
Here we derive the equations for the cosmology in the presence of a bulk scalar field background that depends on $y$ and $t$.  The $ty$ Einstein equation now reads
\beq
\frac{\dot{a}}{a} \frac{n'}{n} - \frac{\dot{a}'}{a} = \frac{\kappa^2}{3} \dot{\phi} \phi'.
\eeq
With a particular scalar field solution, this can in principle be integrated to give
\beq
n = \frac{\dot{a}}{\dot{\bar{a}}} \exp \left[ \frac{\kappa^2}{3} \int \dot{\phi} \phi' \left( \frac{a}{\dot{a}} \right) dy \right].
\eeq
The $tt$ equation no longer generically separates the $a$-function:
\beq
( a^2 )'' - 4 a^2 \left[ -\frac{\kappa^2}{6} V(\phi) - \frac{\kappa^2}{12} \phi'^2 \right] = 2 \dot{\bar{a}}^2 \left( 1- \frac{\kappa^2}{6} \dot{\phi}^2 \left( \frac{a}{\dot{a}} \right)^2 \right) \exp \left[ -\frac{2\kappa^2}{3} \int \dot{\phi} \phi' \left( \frac{a}{\dot{a}} \right) dy \right].
\eeq
In the case of a static scalar field, there are separated solutions, as the function multiplying $a^2$ is time independent, and the inhomogeneous term on the right depends only on time.

The difference of the $tt$ and $yy$ equations gives
\beq
\frac{1}{a^3n} \left[ \partial_t \left( \frac{\dot{a} a^2}{n} \right)-\partial_y \left( a' a^2 n\right) \right] = \frac{2 \kappa^2}{3} V(\phi).
\eeq
\subsection{The IR brane junction conditions}
The last step has to involve matching the IR brane onto this bulk geometry.  The IR brane is embedded as a time-dependent separation, expressed as $y_1 = R(t)$.  We presume that the IR brane potential is simply a tension, $T_1$, which we will take to be de-tuned, in the general case.

To get the junction conditions, we evaluate the jump in the extrinsic curvature on the brane, and match it to the brane localized stress-energy.  We need the induced metric, which is
\beq
h_{MN} = g_{MN}-\eta_M \eta_N,
\eeq
where $\eta$ is the unit normal to the brane.  As the brane has 4-velocity
\beq
u_A = (n \gamma,\vec{0}_3,\beta \gamma),
\eeq
where $\gamma = (1-(\dot{R}/n)^2)^{-1/2}$, $\beta = \dot{R}/n$, it has unit normal
\beq
\eta_A = (n \beta \gamma, \vec{0}_3,\gamma).
\eeq
Now we can go about calculating the extrinsic curvature
\beq
K_{MN} = h_M^L h_N^R \nabla_L \eta_R.
\eeq
Raising the index with the induced metric, we have
\begin{align}
K_0^{~0} &= \frac{ \ddot{R}/n+n'- \dot{R}/n \left( \dot{n}{n}+ 2 \dot{R}{n} n'/n\right)}{n \left( 1- (\dot{R}/n)^2 \right)^{5/2}}, \nonumber \\
K_0^{~5} & = \dot{R} K_0^{~0}, \nonumber \\
K_5^{~0} & =  -K_0^{~5}/n^2, \nonumber \\
K_i^{~j} & =\frac{ \dot{R}/n~ \dot{a}/(a n) + a'/a }{\sqrt{1- (\dot{R}/n)^2}}, \nonumber \\
K_5^{~5} &= - \left( \frac{\dot{R}}{n} \right)^2 K_0^{~0}.
\end{align}
Now the brane stress energy tensor is rather simple in the case of a pure tension:
\beq
S_{MN} = -T_1 h_{MN},
\eeq
and in terms of the tensor $\hat{S}_{MN} = S_{MN} - 1/3 S h_{MN} = 1/3 T_1 h_{MN}$, we can find the first order equation from the $ij$ components of the jump conditions
\beq
[ K_{\mu\nu} ] = -\kappa^2 \hat{S}_{\mu\nu},
\eeq
where the brane tensor $X_{\mu\nu} = e^M_\mu e^N_\nu X_{MN}$, where the $e$'s form a basis for coordinates parametrizing the brane.

The $ij$ condition then gives an equation solving the motion of the brane in our metric background:
\beq\label{eq:1OrderIRJC}
\left. \frac{a'}{a}\right|_{y=R(t)} = \left. \frac{\kappa^2}{6} T_1 \sqrt{1- \beta^2} - \beta \frac{\dot{a}}{an} \right|_{y=R(t)},
\eeq
where metric factors are evaluated at $R_-$ (just to the left of the brane), and we have defined $\beta \equiv \frac{\dot{R}}{n}$.

Another junction condition can also be derived:
\beq\label{eq:2OrderIRJC}
\left. \frac{n'}{n}\right|_{y=R(t)} = \left. \frac{ \frac{\kappa^2}{6} T_1 \left( 1- \beta^2 \right)- \frac{\dot{\beta}}{n}  }{\left( 1 - 2 \beta^2 \right) } \right|_{y=R(t)}.
\eeq
In the case of no bulk matter, this equation gives no additional information, and can be obtained from the first order equation.  However, it is useful to combine these two relations and take the non-relativistic limit, in which case we find a simple and intuitively familiar equation for the evolution of the brane:
\beq
\ddot{R} + \left( 3 \frac{\dot{a}}{a}- \frac{\dot{n}}{n} \right) \dot{R} + n^2 \left[ 3 \frac{a'}{a} + \frac{n'}{n} - \frac{2 \kappa^2}{3} T_1 \right]=0.
\eeq
The second term is a form of 5D Hubble friction for the holographic dilaton, and the last term is an effective potential contribution to its equation of motion.

%%%%%%%%%%%%%%%%%%%%%%%%%%%%%%%%%%%%%%%%%%%%%%%%%%%%%%%%%%%%%%%%%%%%%%%%%%%%%%%%%%%%
%%%%%%%%%%%%%%%%%%%%%%%%%%%%%%%%%%%%%%%%%%%%%%%%%%%%%%%%%%%%%%%%%%%%%%%%%%%%%%%%%%%%

\section{Slow Roll Conditions}
\label{sec:AppendixB}
In this appendix, we provide a detailed derivation of the conditions for the validity of the slow-roll approximation for the time-dependence of the Goldberger-Wise scalar field.

\subsection{UV brane slow roll parameters}
The effective dynamical UV brane mistune is
\beq
\tilde{\delta}_0 \equiv \frac{\kappa^2}{24}\left[ \left( \bar{\phi}'_0 \right)^2  -\left( \phi'_0\right)^2 \right],
\eeq
where $\tilde{\delta}_0$ is now a function of time through a dependence on the location of the IR brane, $R(t)$. If this were changing slowly with time, the cosmology can be approximately described as containing dark radiation, and a slowly changing vacuum energy
\beq
H^2 = \frac{4\bar{\lambda}}{\bar{a}^4}+ \frac{\kappa^2}{12}\left[ \left( \bar{\phi}'_0 \right)^2  -\left( \phi'_0\right)^2 \right].
\eeq
%Now if the right hand side of this equation were constant, the equation could be integrated, giving $H^2 = \frac{4\bar{\lambda}}{\bar{a}^4}+ \frac{\kappa^2}{12}\left[ \left( \bar{\phi}'_0 \right)^2  -\left( \phi'_0\right)^2 \right]$, yielding a solution that is an admixture of radiation and vacuum energy.  While we can't quite do this, we see that if $\Tilde{\delta}_0$ were to be only slowly varying with time, this solution would be accurate up to corrections due to this variation.
%If we can employ such a ``slow-roll'' approximation for the time evolution of the GW scalar field, the effective cosmology seen by a UV localized observer will be well approximated by radiation + slowly evolving vacuum energy.  
Plugging in the approximate solution to the exact equation, we find
\beq
\frac{1}{2} \left[ \dot{H} + 2 H^2 \right] = 2\tilde{\delta}_0 \left( 1 + \frac{\dot{\tilde{\delta}}_0}{4 H \tilde{\delta_0}} \right).
\eeq
We thus have the first of a set of conditions under which we can approximate the evolution of the geometry as one of nearly constant mistunes:
\beq
\epsilon_\text{UV} \equiv  \left| \frac{\dot{\tilde{\delta}}_0}{4 H \tilde{\delta_0}}\right| < 1.
\eeq
In the epochs of radiation domination and vacuum energy domination we have
\beq
\epsilon_\text{UV} \approx \left\{ 
\begin{array}{ll}
\frac{t}{2} \frac{\dot{\tilde{\delta}}_0}{\tilde{\delta}_0} & \text{radiation era} \\
\frac{1}{2} \frac{\dot{\tilde{\delta}}_0}{(2 \tilde{\delta}_0)^{3/2}} & \text{CC era}.
 \end{array} \right.
 \eeq
Since the scalar backreaction is assumed to be small, we can approximate the bulk geometry as the one of constant $\delta_0$, but substituting in the slowly varying $\tilde{\delta}_0$ in its place.

\subsection{IR brane slow roll parameters}

We can quantify slow roll for the effective IR tension by consideration of the boundary conditions.  In the case of constant $\delta_1$ and $\delta_0$, the two IR brane boundary conditions in Eqs.~(\ref{eq:1OrderIRJC}) and (\ref{eq:2OrderIRJC}), are equivalent.  The second can be derived by taking the time derivative of the first, dividing by an effective IR brane Hubble rate, $H_\text{IR} = \frac{\dot{a}}{a} + \frac{a'}{a} \dot{R}$, and then adding the first equation to the result.

Since the effective IR brane mistune $\tilde{\delta}_1$ is \emph{not} constant, this procedure will lead to a deviation of the result of this operation from the second boundary condition by an amount proportional to $\dot{\tilde{\delta}}_1=\tilde{\delta}'_1\dot{R}$.  The residual term (when the equation is normalized by taking the coefficient of $\ddot{R}$ to 1), is given by
\beq
\frac{n^2 \left( 1 - \beta^2 \right)^{3/2}}{H_\text{IR}} \tilde{\delta}'_1 \dot{R}.
\eeq
There are effectively two regimes of this contribution due to the form of $H_\text{IR}$:  $\frac{\dot{a}}{a} > \frac{a'}{a} \dot{R}$ and $\frac{\dot{a}}{a} < \frac{a'}{a} \dot{R}$.

When the first condition is valid, the additional contribution is like an extra term in the friction coefficient of the equation of motion, Eq.~(\ref{eq:VE2ndorder}), which is dominated by the leading $3\frac{\dot{a}}{a}$.  We can thus state that slow roll in this regime is achieved when we satisfy
\beq
\eta_\text{IR} \equiv \left| \frac{n^2 \tilde{\delta}'_1}{3 \left( \dot{a}/a \right)^2} \right| < 1.
\eeq
In the other regime, when $\frac{\dot{a}}{a} < \frac{a'}{a} \dot{R}$, the time derivative of $R$ cancels, and we instead compare with the effective $\frac{\partial V}{\partial R}$ term.

In this case, we must satisfy
\beq
\epsilon_\text{IR} \equiv \left| \frac{\tilde{\delta}'_1}{4 \left( \tilde{\delta}_1+ e^{2R} \tilde{\delta}_0 \right) } \right| < 1.
\eeq
These slow roll conditions are valid throughout a significant regime of the majority of trajectories of interest for our study.  In slow roll inflation, the slow-roll conditions are violated at the exit of inflation, where it is presumed the inflaton quickly finds the minimum of its potential, and its oscillations contribution to reheating.  In our analysis, the portion of the dynamics for trajectories near the minimum of the stabilizing potential is not in the slow roll regime, and we do not expect that using the first order equation with effective constant $\delta_0$ and $\delta_1$ will accurately capture the ``ring-down” of the dilaton, which may or may not include oscillations which we expect would decay to SM particles in realistic scenarios.

\end{document}